\documentstyle[preprint,amsfonts,amssymb,aps,12pt]{revtex}
\input{epsf}
\tighten
\begin{document}
\draft
%
%
\title{Polymer Films in the Normal-Liquid and Supercooled State:\\
A Review of Recent Monte Carlo Simulation Results}
\author{C. Mischler$^1$, J. Baschnagel$^2$%
\footnote{To whom correspondence should be addressed. Email: {\sf baschnag@ics.u-strasbg.fr}}, 
K. Binder$^1$\\[2mm]}
\address{$^1$Institut f\"ur Physik, Johannes-Gutenberg Universit\"at, D-55099 Mainz, Germany}
\address{$^2$Institut Charles Sadron, 6 rue Boussingault, F-67083, Strasbourg Cedex, France}
\maketitle
%
%
%
%
\newcommand{\mr}[1]{{\rm #1}}
\renewcommand{\vec}[1]{\mbox{\boldmath$#1$\unboldmath}}
%
%
%
%
%
\begin{abstract}
The present paper reviews recent attempts to study the development of glassy behavior in thin polymer films by 
means of Monte Carlo simulations. The simulations employ a version of the bond-fluctuation lattice model, in
which the glass transition is driven by the competition between an increase of the local volume requirement of 
a bond, caused by a stiffening of the polymer backbone, and the dense packing of the chains in the melt. The
melt is geometrically confined between two impenetrable walls separated by distances that range from once to about
fifteen times the bulk radius of gyration. The confinement influences static and dynamic properties
of the films: Chains close to the walls preferentially orient parallel to it. This orientation tendency 
propagates through the film and leads to a layer structure at low temperatures and small thicknesses.
The layer structure strongly suppresses out-of-plane reorientations of the chains. In-plane reorientations
have to take place in a high density environment which gives rise to an increase of the 
corresponding relaxation times. On the other hand, local density fluctuations are enhanced if the film thickness 
and the temperature decrease. This implies a reduction of the glass transition temperature with decreasing film
thickness.
\end{abstract}
%
%
\pacs{{\sf PACS}: 61.20.Ja,61.25.Hq,64.70.Pf\\%
accepted for publication by {\em Adv.\ Colloid Interf.\ Sci.}
}
%
%
\section{Introduction and Overview}
\label{intro}
Many recent studies deal with the influence of confinement on the dynamic behavior of glass forming liquids
\cite{vigo1997,confit2000}. Besides the practical importance of such systems (polymer films as protective
coatings, flow through porous materials, etc.) one motivation for this research is that it might also provide 
a better understanding of the glass transition in the bulk \cite{vigo1997,ictp1999,goetze4,jaeckle}. If the 
glass transition was driven by an underlying correlation length $\xi$ which grows with progressive supercooling 
towards the transition temperature $T_\mr{g}$, confinement should truncate the growth as soon as $\xi$ becomes 
comparable to the size, $L$, of the restrictive geometry. Marked deviations from the bulk behavior are then expected. 

However, the trend of these deviations is not obvious. Should the dynamics be accelerated or slowed down? 
Theoretical arguments may be put forward for both scenarios. For instance, one could speculate that 
the glass transition is caused by some subtle kind of static order which characterizes the spatial arrangement
of the particles in the glassy phase and spontaneously develops during supercooling. The correlation length
should then measure the growth of these (solid-like) regions. If this was true, one could invoke an
analogy to the theory of spin glasses \cite{spinglass} and suggest that the structural relaxation time 
$\tau$ scales as $\tau \sim \xi^z$ ($z$ being a dynamic critical exponent). When $\xi \simeq L$, $\tau$ 
should level off and the dynamics should become faster in confined geometry. On the other hand, one can also
adopt the point of view that supercooling makes the dynamics of a glass former become progressively cooperative.
Cooperativity means that the displacement of a particle is predicated upon the simultaneous rearrangement of many
neighboring particles and the correlation length measures the spatial extent for these cooperative
processes. There is evidence for such a dynamic heterogeneity from recent simulations 
\cite{FyneHarro_ictp1999,YamamotoOnuki1998,DoliwaHeuer,gns,dgpkp,bdbg} and experiments 
\cite{kb,wclsw,silly_rev,Ediger_AnnRevPhysChem2000}. 
In spatial confinement some of these processes should be suppressed. Therefore, one expects $\tau$ to increase 
and the dynamics to slow down compared to the unrestricted bulk.

A variety of experiments 
\cite{confit2000,JacksonMcKenna1996,WendtRichert1999,Schuelleretal1995,Kremeretal1999,Barutetal1998,Keddieetal1994,%
forrest1,forrest2,forrest3,forrest4,Anastasiadisetal2000,FNdePablo}, computer simulations 
\cite{confit2000,bt1999,MansfieldTheo1991,nl1999,fl1995,YamamotoKim2000,skb,Galloetal2000,compodyn,ssg2000} 
and theoretical approaches \cite{confit2000,confit_jaeckle,PGG2000,BarratBocquet1995} 
have attempted to display the phenomenology and to elucidate the underlying mechanisms of dynamics in confinement. 
The systems studied range from simple liquids \cite{bt1999,nl1999,fl1995,YamamotoKim2000,skb,BarratBocquet1995}, 
over molecular and hydrogen-bonded liquids 
\cite{JacksonMcKenna1996,WendtRichert1999,Schuelleretal1995,Kremeretal1999,Barutetal1998,Galloetal2000} to polymers 
\cite{Keddieetal1994,forrest1,forrest2,forrest3,forrest4,Anastasiadisetal2000,FNdePablo,compodyn,ssg2000,PGG2000}. 
The geometries considered involve three-dimensional cavities \cite{Barutetal1998,nl1999,confit_jaeckle}, pores 
\cite{JacksonMcKenna1996,WendtRichert1999,Schuelleretal1995,Kremeretal1999,skb,Galloetal2000}, nano-sized fillers
embedded in polymer melts \cite{ssg2000}, 
and thin films \cite{Keddieetal1994,forrest1,forrest2,forrest3,forrest4,Anastasiadisetal2000,FNdePablo,bt1999,%
MansfieldTheo1991,fl1995,YamamotoKim2000,compodyn,PGG2000,BarratBocquet1995}, exhibiting different
interactions between the glass former and the walls of the confinement. These interactions and the arrangement of 
the molecules close to the walls introduce boundary effects which perturb the bulk structure. These effects can
reinforce or mask the influence of pure confinement on the dynamics. Therefore, a good knowledge and control of 
the surface properties is important for the interpretation of the results. 

Such knowledge and control is typically available in computer simulations. Simulations study precisely defined -- though
often highly idealized -- systems to reveal possible consequences of various kinds of confinement. A special kind of 
confinement which may only be realized in computer simulations is a size variation of the simulation box while
maintaining periodic boundary conditions. This means the following: Usually, the simulated system is contained
in a cubic box which is replicated in all spatial dimensions. If a particle leaves the box on one side, an identical
image particle simultaneously enters the box from the opposite side. The simulation cell can thus be thought of as a 
section of a macroscopic system. The fact that the adjacent boxes are identical copies does not influence the physical
properties of the system if the box size is large (and if there are no such complications, such as long-range
interactions, spontaneous ordering close to phase transitions, etc.). However, if it shrinks, finite-size
effects may occur. The development of such finite-size effects during supercooling has been observed in simulations 
of binary Lennard-Jones mixtures \cite{BuechnerHeuer1999}, of hard sphere mixtures \cite{KimYamamoto2000,JaeckleKawai2000} 
and of silica 
\cite{Horbachetal1,Horbachetal2}, and was also treated analytically \cite{confit_jaeckle}. Generally, one finds the 
dynamics to be slowed down with shrinking box size (see \cite{BuechnerHeuer1999} for a comparative discussion of the 
results from Refs.~\cite{BuechnerHeuer1999} and \cite{KimYamamoto2000}).

An important property of these studies is that the confinement is weak. There is no specific interaction with the
boundary and no spatial arrangement close to the surface of the simulation box. Changes of the bulk structure are very
mild or not present at all. The same situation is also realized in recent simulations of the relaxation
dynamics in spherical cavities \cite{nl1999} and pores \cite{skb,confit_skb}. In these studies, a sphere or a cylinder 
is cut out of a large bulk sample. Subsequently, only the inner particles are allowed to move, whereas the outer particles
are frozen. They realize the walls of the confinement. Since the walls are rough and adapted to the liquid
structure, a fluid particle can be trapped in cavities, into which it perfectly fits. Relaxation out of these traps
is thus sluggish so that one finds a slowing down of the dynamics close to the walls. This slow relaxation also 
retards the dynamics of particles in the inner part of the liquid \cite{skb,confit_skb}.

While the impact of confinement on the structure is negligible in these studies, wall-induced perturbations of the bulk 
structure occur in simulations of thin film geometries. If the film geometry is realized by two free surfaces 
(freely standing films), the density of the liquid decreases from the bulk value to zero across the liquid/vacuum 
interface \cite{bt1999,MansfieldTheo1991,DorukerMattice1999}. The interface sharpens with decreasing temperature. 
On the other hand, if the glass former is embedded between two impenetrable and smooth walls, there are pronounced 
density oscillations starting with a value larger than the bulk density at the wall. These oscillations can decay towards 
the bulk density with increasing distance from the wall or propagate through the film, depending on the film thickness and 
temperature considered. These structural differences also affect the dynamic behavior. For the freely standing films one 
generally finds that the particles closer to the interface are more mobile than inner particles which behaves bulk-like
\cite{bt1999,MansfieldTheo1991,DorukerMattice1999}. Therefore, the overall dynamics of the film is faster than the bulk 
at the same temperature. This should lead to a decrease of the average glass transition temperature. The same 
interpretation is also suggested by an analysis of recent experimental results for freely standing polystyrene films
\cite{forrest4} and by comparitive computer simulations of freely standing and supported polymer film models
\cite{TNdePablo}. 

Whereas the observed speeding up of the particle motion close to a free interface is intuitively expected, the situation 
is not so clear for glass formers confined between two completely smooth, impenetrable walls which exert no preferential 
attraction on the particles, but merely represent a geometric confinement. Simulations of glassy polymer films 
\cite{confit_bmb,confit_vbb} suggests an acceleration of the local relaxation dynamics, whereas simulations of binary hard 
sphere mixtures rather find a slowing down \cite{fl1995,YamamotoKim2000}. A tentative explanation of this difference 
could be as follows: The aforementioned density oscillations are much more pronounced for the binary mixtures than for
the polymer films. This is a consequence of the model parameters used in the simulations. For the mixtures there seem to 
be densely filled layers of particles stacked on top of each other at low temperature. Motion has to occur in this very 
high density environment which might therefore be slower than in the bulk and in the simulations of the polymer films. 

Thin film geometries are also extensively studied experimentally, especially for poly\-sty\-rene melts 
\cite{Keddieetal1994,forrest1,forrest2,forrest3,forrest4,Anastasiadisetal2000,FNdePablo}. The 
geometries investigated range from freely standing films \cite{forrest1,forrest2,forrest3,forrest4} to supported films 
(two inequivalent surfaces: ${\rm SiO_2}$ and air) \cite{Keddieetal1994,forrest2,forrest3,FNdePablo} and polystyrene 
embedded 
between two ${\rm SiO_2}$ substrates, realized either by capping the film with another silicon-oxide layer \cite{forrest2} 
or by intercalating the polymer in layered-silicate hosts \cite{Anastasiadisetal2000}. A systematic variation of film 
thickness and chain length has revealed many interesting results which have been summarized in recent comprehensive 
reviews \cite{ForrestJones2000,ForrestDalnoki-Veress2000} and elicited theoretical explanations \cite{PGG2000}.

The present paper also deals with the glass transition of a thin polymer film. It presents Monte Carlo simulation results
for a simple model of a non-entangled polymer melt confined between two solid walls. Its main focus is to investigate
static and dynamic properties when polymer films of different thicknesses are progressively supercooled. The paper is 
organized as follows: Section~\ref{model} describes the background of the model and the employed simulation technique. 
Sections~\ref{static} and \ref{dynamic} illustrate the influence of spatial confinement on the structure of the melt and 
the resulting consequences for the dynamic behavior. The last section~\ref{sum} summarizes and discusses the results.

\section{Coarse-Grained Lattice Simulations of Polymer Films}
\label{model}
The present approach uses a lattice model: the bond-fluctuation model \cite{bfl1,bfl2,bfl3}. This model
is intermediate between a highly flexible continuum treatment and standard lattice models of polymers 
\cite{as_kbrev,kb_kbrev,bi_kr}. With the latter it has in common the simple lattice structure which
is very efficient from a computational point of view \cite{wp3}. However, it differs from them in that the
set of bond vectors is not limited by the coordination number of the lattice, but much larger. In this 
respect, it resembles more a continuum model. 

A monomer of 
the bond-fluctuation model does not directly correspond to a chemical monomer. It should rather be thought
of as representing a group of chemical monomers (comprising typically $3$ to $5$ monomers for simple 
polymers, such as polyethylene \cite{kb_kbrev,cg_review}). A lattice bond should thus be interpreted 
as the vector joining these groups. This coarse-grained vector can fluctuate in length and direction to a
much larger extent than its chemical counterpart. The bond-fluctuation model accounts for this flexibility
of the coarse-grained bond vector by associating a monomer not with a single lattice site, but with a unit 
cell of a simple cubic lattice. The monomers can be connected by 108 different bond vectors which are 
chosen such that local self-avoidance of the monomers and uncrossability of the bond vectors during the 
simulation are guaranteed.

\vspace{4mm}

\noindent {\bf Energy Function and Geometric Frustration.}
In addition to excluded volume interaction and chain connectivity an energy function ${\cal H}(\vec{b})$ is 
introduced for the 
bond vectors $\vec{b}$. It favors bonds of length $b=3$ and directions along the lattice axes 
(${\cal H}(\vec{b})=0$) in comparison to the rest of available bond vectors (${\cal H}(\vec{b})
=\epsilon$) \cite{acs98,kb_rev93,cool1}. 
Figure~\ref{coarse+hard} illustrates the effect of this two-level Hamiltonian. If temperature decreases, each 
bond attempts to adopt the ground state ${\cal H}(\vec{b})=0$. A bond in the ground state blocks four lattice 
sites which can no longer be occupied by any other monomers due to the excluded volume interaction. This 
reduces the amount of accessible volume and generates a competition between the internal energy of a bond 
and the local arrangement of other monomers around it. This competition forces some bonds to remain in the excited 
state. They are geometrically frustrated \cite{kb_rev93,cool1}. For instance, the bond of the lower chain in 
Fig.~\ref{coarse+hard} could only reach the ground state if the monomer of the upper chain moved away. However, 
this motion is only possible if the constraints acting on this monomer are released which might in turn require 
the motion of further distant monomers. 
The relaxation of local geometric frustration is thus predicated upon 
the cooperative rearrangement of many monomers, which can become very sluggish due to mutual blocking
at low temperatures. Therefore, the development of the geometric frustration during the cooling 
process causes the glassy behavior of the model. It is also the driving force which the Gibbs-Di Marzio 
theory makes responsible for the glass transition of polymer melts \cite{gibbs2,comp_po_sci,matt_entropy}.

\vspace{4mm}

\noindent {\bf Monte Carlo Methods: Real versus Artificial Dynamics.}
The competition between energetic and packing constraints leads to a slow, glass-like 
structural relaxation of the melt if the usual bond-fluctuation dynamics is used. This dynamics consists 
of the following steps: First, a monomer and a lattice direction are chosen at random. Then, it is checked whether 
the targeted lattice sites are empty (excluded volume interaction) and whether the new bonds belong to the allowed 
set (maintenance of chain connectivity). If these conditions are satisfied, the energy change $\Delta E$ associated 
with a monomer displacement in the chosen lattice direction is calculated. The attempted move is
accepted with probability $\min\left(1,\exp\left[-\beta\Delta E\right]\right)$ (Metropolis criterion \cite{bi_kr,bi_he}), 
where $\beta=1/k_\mr{B}T$ is the reciprocal temperature (the temperature is measured in units of $\epsilon/k_\mr{B}
$). These local moves are supposed to mimic a random force exerted on a monomer by its environment. They lead 
to Rouse-like dynamics which is typical of short polymers in dense melts \cite{doi,dsk,bipa_rev}. 

The local dynamics gives rise to a strongly protracted structural relaxation of the melt. Although
this is exactly the physical phenomenon that we are interested in, such a slow relaxation is very disadvantageous
to quickly equilibrate the system during the cooling process. To circumvent this problem one can exploit
the property of the Monte Carlo technique that the elementary move may be adapted at will. Since the final 
equilibrium state is independent of the way by which it was reached, the realistic local dynamics may be replaced by 
an artificial one which uses non-local moves. A non-local move involves many (or even all) monomers along the 
backbone of a chain. 

The so-called ``slithering snake dynamics'' is an example for such a collective move
\cite{as_kbrev,bi_kr}. In this dynamic scheme, one tries to attach a bond vector to one of the ends of a polymer. 
Both the vector and the end monomer are randomly chosen. If the attempt does not violate the excluded volume
restriction, the move is accepted again with probability $\exp [-\Delta E/k_\mr{B}T]$. But now, $\Delta E$ represents
the energy difference between the newly added bond and the last bond of the other end of the chain, which is removed
when the attempt is accepted.
Whereas the local dynamics propagates the chains by one lattice constant for every accepted move, the 
slithering-snake dynamics shifts the whole chain by one lattice constant. Therefore, one expects this
algorithm to be faster by a factor of the order of the chain length already at high temperatures. The simulations
show that this expectation is borne out and furthermore that the algorithm is very efficient in equilibrating low 
temperatures \cite{matt1,volker1}.  

\vspace{4mm}

\noindent {\bf Simulation Parameters.} 
In the present study a chain always consists of $N=10$ monomers. When taking into account that a lattice monomer roughly 
corresponds to a group of three to five chemical monomers (see above) our simulation deals with fairly short, 
non-entangled \cite{wp3,mwb,kbmb} oligomeric chains. The number of chains, $K$, in the rectangular simulation box 
(of size $L \times L \times D$)
is adjusted so that the volume fraction of occupied lattice sites is $\phi=8NK/DL^2= 0.5\bar{3}$. 
This value is a compromise between two requirements: it is high enough so that the model realizes the typical 
properties of dense melts \cite{wp3}, and low enough to allow a sufficient acceptance rate of chain moves to 
equilibrate the melt by slithering-snake dynamics \cite{matt1}. 

The dimension of the simulation box parallel to the walls is $L=60$ (in units of the lattice constant; see
Fig.~\ref{coarse+hard}). Since the maximum end-to-end
distance is $R_\mr{max}=27$ ($=9 \times 3=(N-1) \times \mbox{bond length in the ground state}$), we have
$L/2 > R_\mr{max}$ so that there are no finite-size effects (this has been tested explicitly by 
simulations with $L=120$). Impenetrable, completely smooth and structureless walls are situated at $z=1$ and $z=D$.
Technically, these walls are generated by disallowing monomer moves to $z<1$ or $z>D$.
The film thickness ranges from $D=6$ ($\approx 1.5 R_\mr{g}$; $R_\mr{g}=$ bulk radius of gyration) to $D=60$
($\approx 15 R_\mr{g}$). 

In order to improve the statistics several independent simulation boxes are treated in parallel. The total 
statistical effort involves between $43200$ and $86400$ monomers, depending on the system under consideration.
This allows us to keep the numerical uncertainties of the results at a low level. Unless otherwise stated,
the statistical errors are always of the order of the size of the symbols in the following figures.

\section{Static Properties: Gyration Tensor and Density Profiles}
\label{static}
Structural properties represent an important input for analyzing the dynamics of a polymer melt (and of any
other system in general). Close to a solid interface the structure of the melt markedly deviates from the behavior 
of the unconstrained bulk \cite{fleer,sanchez}. This is pointed out by analytical approaches 
\cite{hmc,hpmcw,mg,yet1998,curro,pkryd,ykhs,freed,wy,wspcg,kr,theo3}, 
computer simulations of lattice \cite{bitbri,bah,pakula,wang1,composit,feature,pdp} and continuum models 
\cite{ssg2000,hpmcw,mg,curro,pkryd,ykhs,yoon_rev93,bithad,wmy,vaca1,kumar1,ron1,yet}, and recent experiments 
\cite{kmkss} (however, see also \cite{jkhbr} for different experimental results). 
The present section describes the influence of hard walls on the static properties of our model.

\subsection{Gyration tensor}
A polymer chain in the bulk can adopt a multitude of different configurations. Each configuration has a certain
distribution of monomers around its center of mass. This distribution determines the instantaneous shape of the 
chain.
Classical mechanics suggests that a quantitative measure of this distribution may be provided by the moment of 
inertia tensor $\mathsf{\Theta}$ \cite{s2}
\[
\mathsf{\Theta} = \big(\mr{Tr}\, \mathsf{Q}\big)\mathsf{1} - \mathsf{Q} \; ,
\]
where {\sf 1} denotes a $3\times 3$ unit matrix and $\mr{Tr}\, \mbox{\sf Q}$ is the trace of the tensorial 
generalization of the radius of gyration (note that $\langle \mr{Tr}\,{\mathsf{Q}}\rangle =R^2_\mr{g}$)
\begin{equation}
Q_{\alpha\beta}=\frac{1}{N}\sum_{n=1}^N\Big[\left(r_{n,\alpha}-R_{{\rm cm},\alpha}\right)
\left(r_{n,\beta}-R_{{\rm cm},\beta}\right)\Big] \qquad (\alpha,\beta = 1,2,3) \; .
\end{equation}
Here, $r_{n,\alpha}$ and $R_{{\rm cm},\alpha}$ are the $\alpha$th spatial component of the position vectors to
the monomer $n$ ($\vec{r}_n$) and to the center of mass ($\vec{R}_\mr{cm}$), respectively. 

The average of 
$\mathsf{\Theta}$ or equivalently of the gyration tensor $\mathsf{Q}$ are good indicators of the characteristic
shape of a polymer because they reflect the average distribution of monomers in the internal coordinate system
of a chain. This coordinate system is given by the principal axes obtained after diagonalization of $\mathsf{Q}$.
The corresponding eigenvalues, $\lambda_1$, $\lambda_2$ and $\lambda_3$, measure the length of the axes. If the 
(instantaneous) monomer distribution was spherical, all eigenvalues would be equal. 

However, the seminal work of \v{S}olc and Stockmayer already pointed out that the average shape of a (single
isolated) polymer is far from spherical \cite{SolcStock}. This conclusion has been corroborated by various further studies 
\cite{s2,cag,jtc,em}. All principal axes of a polymer have 
different lengths. For a random walk the ratio of the average eigenvalues is given by: 
$\langle \lambda_1\rangle : \langle \lambda_2\rangle : \langle \lambda_3\rangle = 12.07 : 2.72 : 1$ \cite{em}. 
Therefore, even a random walk is distorted with respect to a sphere: Its size is extended along the 
largest principal axis and reduced in directions of the axes corresponding to $\langle \lambda_2\rangle$ and 
$\langle \lambda_3\rangle$. Since a random walk is a viable model for (the large scale properties of) a chain in 
the melt, the average shape of chains in the melt rather resembles a flattened ellipsoid than a sphere%
\footnote{Recent studies \cite{jtc,em} of the shape of Gaussian chains indicate that the visualisation of a 
random-walk polymer as a flattened ellipsoid is not completely correct. The density distribution of 
monomers in the coordinate system of the principal axes exhibits a slight minimum at the origin (i.e., at the 
center of mass) for the largest axis, whereas it has a maximum for the other two axes. Therefore, the shape is 
rather dumbbell-like.}.

In addition to the eigenvalues two other quantities were introduced to discuss deviations from a spherical
structure: the asphericity $\Delta_0$ and the prolateness $S_0$ \cite{s2,rg}. They are defined by
\begin{equation}
\Delta_0=\frac{3}{2}\,\frac{{\rm Tr}\,\mbox{\sf Q}^2}{\big( {\rm Tr}\,\mbox{\sf Q}\big)^2}
=\frac{1}{6}\,\frac{\sum\limits_{\alpha=1}^3\left(\lambda_{\alpha}-\overline{\lambda}\right)^2}{\overline{\lambda}^2}
\label{defDelta}
\end{equation}
and
\begin{equation}
S_0 = 27\,\frac{\det \mbox{\sf Q}}{\big({\rm Tr}\, \mbox{\sf Q}\big)^3}
=\frac{\prod\limits_{\alpha=1}^3\left(\lambda_{\alpha}-\overline{\lambda}\right)}{\overline{\lambda}^3}
\label{defS}
\end{equation}
with $\overline{\lambda}=(\lambda_1+\lambda_2+\lambda_3)/3$.

For a sphere one has $\lambda_\alpha=\overline{\lambda}$ so that $\Delta_0 = S_0 =0$. Two extreme deviations
from this high symmetry can be considered: On the one hand, the sphere may be squashed to a disk. Then, 
$\lambda_1=0$ and $\lambda_2=\lambda_3=3 \overline{\lambda}/2$ which yields $\Delta_0=1/4$ and $S_0 = -1/4$. 
On the other hand, it can be stretched to a rod. Then, $\lambda_1=3\overline{\lambda}$, whereas $\lambda_2=
\lambda_3=0$ so 
that $\Delta_0=1$ and $S_0 = 2$. Therefore, the asphericity only measures deviations from the spherical
structure, whereas the prolateness additionally determines by its sign whether the deviation is disk-like, 
i.e., ``oblate'' ($S_0 < 0$), or elongated, i.e., ``prolate'' ($S_0 > 0$).

\subsection{Asphericity and Prolateness of Polymer Films}
\noindent {\bf High-Temperature Results.}
Figures~\ref{aspher.allT.allD} and \ref{prolat.allT.allD} compare the variation of $\langle \Delta_0(z_\mr{cm})
\rangle$ and $\langle S_0(z_\mr{cm}) \rangle$ with the distance of the chain's center of mass, $z_\mr{cm}$, from
the (left) wall for different film thicknesses $D$ at $T=\infty$. Here, $\langle \bullet \rangle$ denotes the 
average over all chains in the system. For $T=\infty$ only excluded volume interactions between the monomers and
between the monomers and the walls are effective. This temperature is therefore representative of the 
high-temperature liquid state of the melt. In this state, the average shape of chains in the inner part of the 
film resembles a prolate ellipsoid. The numerical values for $\langle \Delta_0(z_\mr{cm})\rangle$ and $\langle 
S_0(z_\mr{cm})\rangle$ are close to those of a random walk \cite{cag}. The bulk-like inner part extends from
the middle of film to about $z_\mr{cm} \approx 2 R_\mr{g} = 7.36$, where the interfacial region starts. Since
the melt is confined between two walls, a bulk-like inner part can only be observed if $D \geq 18$ ($\approx 5 
R_\mr{g}$). For smaller thicknesses ($D\leq 12\approx 3 R_\mr{g}$) the film just consists of interfacial region 
because the perturbations of the structure, which propagate from both walls, interfere. Similar results have
already been observed before \cite{bah,pakula,pdp,kumar1,yet,yethall}.

Let us first consider the larger films $D \geq 18$. For these thicknesses the profiles of $\langle \Delta_0
(z_\mr{cm}) \rangle$ and $\langle S_0(z_\mr{cm}) \rangle$ are independent of $D$. When $z_\mr{cm} \lesssim 2 
R_\mr{g}$, the chains first slightly expand along the longest principal 
axis, then shrink, pass through a minimum and finally become very elongated when their center of mass lies at the 
wall (i.e., at $z=1$). This variation of the eigenvalues is accompanied by a reorientation of the principal axes
(see \cite{psitf}). Whereas the chains can orient freely in the bulk-like inner region of the film and thus
appear spherical on average, the solid wall singles out those configurations, in which the two largest axes are
aligned parallel to it. The corresponding eigenvalues are bigger than those measured in the film center
[$\langle \lambda_1(z_\mr{cm}=1) \rangle \simeq 12.81 > \langle \lambda_1(z_\mr{cm}=15)\rangle \simeq 10.23$, 
$\langle \lambda_2(1)\rangle \simeq 2.40 > \langle \lambda_2(15)\rangle \simeq 2.32$], 
whereas the third eigenvalue is considerably smaller 
[$\langle \lambda_3(1) \rangle \simeq 0.34 < \langle \lambda_3(15) \rangle \simeq 0.80$]. 
Therefore, the chains are not only oriented parallel to the wall, but also distorted. They adopt a rather 
flattened shape. This influence of the wall on the structure of the chains is also confirmed by other simulations
\cite{mg,curro,pkryd,bitbri,bah,pakula,wang1,pdp,bithad,kumar1,yet,yethall}. 

A distortion of the structure and orientation of the chains is entropically unfavorable because they strongly restrict
the number of accessible configurations. As $z_\mr{cm}$ increases, the disk-like ellipsoid therefore turns away 
from the parallel alignment, shrinks in directions of the longest axes and expands along the smallest
axis. This effect is most pronounced for $z_\mr{cm}=3$, where the polymer concentration has a maximum 
(see Fig.~\ref{chaindensity.allT.allD}). In order to accomodate many chains at the same distance each chain has
to be compressed so that the average shape is least prolate [$\langle \lambda_1(z_\mr{cm}=3) \rangle 
\simeq 8.86$, $\langle \lambda_2(z_\mr{cm}=3) \rangle \simeq 2.31$, $\langle \lambda_3(z_\mr{cm}=3) \rangle 
\simeq 0.84$]. This effect is stronger for $D=6$ because the chain density is much higher at $z_\mr{cm}=3$
(middle of this film) than for the thicker film (see Fig.~\ref{chaindensity.allT.allD}).

If $z_\mr{cm}$ increases further, the chain expands again and $\langle \Delta_0(z_\mr{cm}) \rangle$ and $\langle 
S_0(z_\mr{cm}) \rangle$ go through weak maximum at $z_\mr{cm}=6$ before crossing over to the respective bulk
values. The maximum corresponds to chains which are on average slightly oriented in direction perpendicular to the 
wall. Such chains can (presumably) already touch the wall with some of their monomers and thus contribute to the 
large monomer concentration at $z=1$ (see Fig.~\ref{moden}). This interpretation is corroborated by simulations
of liquid $n$-tridecane, in which the variation of the monomer distribution around the center of mass shows exactly
this behavior, as $z_\mr{cm}$ approaches the wall \cite{vaca3}. Since $z_\mr{cm}=6=D/2$ for $D=12$, chains at 
$z_\mr{cm}=6$ can reach both walls, which enhances $\langle \Delta_0(z_\mr{cm}) \rangle$ and $\langle S_0
(z_\mr{cm}) \rangle$ for this particular thickness in comparison to the other films.

A similar behavior can also be observed for quantities which are experimentally better accessible than 
$\langle \Delta_0(z_\mr{cm}) \rangle$ or $\langle S_0(z_\mr{cm}) \rangle$. For instance, for the radius of 
gyration \cite{composit,feature} (or the end-to-end distance \cite{mrs_98}). When measuring the components of radius 
of gyration parallel and perpendicular to the wall, one finds that the parallel component $R_{\mr{g},\|}(z_\mr{cm})$ 
is large at $z_\mr{cm}=1$, whereas $R_{\mr{g},\perp}(z_\mr{cm}=1)$ is small (see Fig.~\ref{rgden}). With increasing 
separation from the wall the
perpendicular component $R_{\mr{g},\perp}(z_\mr{cm})$ increases towards a pronounced maximum at $z_\mr{cm}=6$ 
before crossing over to the bulk value. On the other hand, $R_{\mr{g},\|}(z_\mr{cm})$ has a shallow minimum 
at $z_\mr{cm}=6$ only, but otherwise decreases continuously towards the bulk value. For both components the bulk 
value is reach if $z_\mr{cm} \approx 2 R_\mr{g}$. This behavior is found in many other simulations
\cite{mg,curro,pkryd,bitbri,bah,pakula,wang1,pdp,bithad,kumar1,yet,yethall}, can be reproduced by self-consistent
field theory (for confined binary polymer blends) \cite{rike} and is also suggested by some experiments
\cite{kmkss} (however, \cite{jkhbr} reports that $R_\mr{g}$ remains essentially bulk-like, even for $D/R_\mr{g} 
\approx 0.5$).

\vspace{4mm}

\noindent {\bf Dependence on Temperature.}
The insets of Figs.~\ref{aspher.allT.allD} and \ref{prolat.allT.allD} illustrate the temperature dependence of
$\langle \Delta_0(z_\mr{cm}) \rangle$ and $\langle S_0(z_\mr{cm}) \rangle$ for the smallest film thickness
$D=6$ ($\approx 1.5R_\mr{g}$). Although this film has no bulk-like inner part, the infinite temperature
results for the asphericity and prolateness at $z_\mr{cm}=1$ coincide with those obtained for larger $D$.
Therefore, the structure of the chains at the wall is independent of film thickness. Other simulations support 
this finding \cite{bah,yet}. When $z_\mr{cm}$ increases,
$\langle \Delta_0(z_\mr{cm}) \rangle$ and $\langle S_0(z_\mr{cm}) \rangle$ decrease and become much smaller
than the bulk value in the middle of the film. This is (probably) a consequence of the high polymer concentration 
at $z_\mr{cm}=3$, as pointed out above (see also Fig.~\ref{chaindensity.allT.allD}). 

If the film is progressively supercooled, this behavior changes in two respects: First, the values of 
$\langle \Delta_0(z_\mr{cm}) \rangle$ and $\langle S_0(z_\mr{cm}) \rangle$ at the wall increase. This indicates 
that the chains become more prolate with decreasing temperature. Second, the minimum at $z_\mr{cm}=3$ turns into
a maximum of about the same height as found at the walls. Furthermore, there are minima between the maxima at 
$z_\mr{cm}=1,3,5$ which have almost the same value as that at the walls for $T=\infty$. The resulting 
zig-zag structure is in phase with the profile of the chain density (Fig.~\ref{chaindensity.allT.allD}). 
Contrary to the behavior at $T=\infty$, chains located at positions of high density are extremely prolate. This
means that all chains of the film, and not only those at $z_\mr{cm}=1$, are preferentially oriented parallel to the 
walls and flattened. 

This temperature dependence of the profiles is a consequence of the model's energy function. Remember that this 
function favors large bond vectors which point along the lattice directions and are thus parallel to the walls. 
At low temperature, the energy function will therefore reinforce the influence of the walls to align chains 
parallel. The combined effect should be to 
single out those chain configurations which almost lie completely in the lattice layer next to the wall. Since
this layer is rather densely filled (see discussion of Fig.~\ref{chaindensity.allT.allD}) and thus essentially
impenetrable, it represents another (rugged) wall in front of the original one. Effectively, the film has 
become thinner by two lattice constants (one from each wall). This reduction of the available volume limits the 
orientational freedom of the remaining chains in the middle of the film so that they also align parallelly.

\subsection{Density Profiles of Chains and Monomers}

\noindent
{\bf Chain Profiles.}
The discusssion of the previous section pointed out that the shape of chains at the wall is flattened.
Such a deviation from the bulk structure and their parallel orientation are entropically unfavorable. At $T=\infty$,
where only entropic effects are present, one can therefore expect the polymers to avoid the vicinity of the walls.
Figure~\ref{chaindensity.allT.allD} shows that the chain concentration, $\rho_\mr{c}(z_\mr{cm})$,
is in fact negligible at $z_\mr{cm}=1$ for all film thicknesses. On the other hand, if $z_\mr{cm}$ increases, the 
concentration quickly rises, attains a maximum close to $z_\mr{cm} = R_\mr{g}$ and then decreases towards the bulk 
density which is reach for $z_\mr{cm} \gtrsim 2 R_\mr{g}$. This behavior is found for $D \geq 12$ ($\approx 3 R_\mr{g}$)
and confirmed by simulations with other models 
\cite{mg,curro,pkryd,bitbri,bah,pakula,wang1,pdp,bithad,kumar1,yet,yethall}. 
The influence of film thickness on $\rho_\mr{c}$ is 
weaker than for the asphericity and prolateness discussed in the previous section. Large deviations are only seen 
for the thinnest film $D=6$, where $D$ is so small that the depletion effects occurring at both walls interfere 
and lead to a high concentration in the middle at $z_\mr{cm}=3$. The same argument was also put forward to interprete
the simulation results of \cite{yet}.

If temperature is reduced, the shape of $\rho_\mr{c}(z_\mr{cm})$ alters completely.
This is exemplified again for $D=6$ in the inset of Fig.~\ref{chaindensity.allT.allD}. 
Since a decrease of temperature favors long bond vectors along the lattice axes, the triangle-shaped profile of 
$T=\infty$ with a low concentration at the wall gradually turns into an enrichement of chains at $z_\mr{cm} = 1$.
These chains lie almost flat in the lattice layer next to the wall. The maximum number of chains which a layer
can accomodate at low temperatures can be estimated as follows: Let us assume that a chain is a rod with 
$N-1=9$ bonds in the ground state (bond length $=3$). The end-to-end distance is then 27. Due to excluded
volume interactions, a $60 \times 60$ lattice layer can take up 60 chains at maximum. So, $\rho_\mr{c}^\mr{max}
(z_\mr{cm}=1)= 0.01\bar{6}$. A comparison of this estimate with  Fig.~\ref{chaindensity.allT.allD} shows that the 
layer at the wall is very dense. Due to excluded volume interaction the following layer is depleted,
which in turn allows for a larger concentration in the subsequent layer and so on. This zig-zag 
structure is a result of the interplay between the model's energy function, the wall and the underlying lattice.

\vspace{4mm}
\noindent {\bf Monomer Profiles.}
The monomer profile $\rho_\mr{m}(z)$ is defined as the average density of monomers which are situated at a 
distance $z$ from the (left) wall%
\footnote{The position of a monomer is associated with the $z$-value of the lower left corner of its unit cell.}.
It exhibits an oscillatory structure not only at low, but also at high
temperatures (see Fig.~\ref{moden}). This structure can be explained as a result of the competition between 
packing constraints and 
loss in entropy. The loss in entropy is caused by the reduction of accessible chain configurations near an
impenetrable wall. It produces an effective repulsive force pointing away from the wall. This force competes
with another effective force exerted by the densely packed chains in the inner part of the film. They tend to 
push the polymers, which are close to the wall, towards the wall. At melt-like densities packing constraints 
dominate and lead to an enrichment of monomers at the wall. Due to the mutual exclusion the monomer
concentration is reduced in the next layer, which in turn allows an enrichment in the subsequent layer
and so on. Therefore, the monomer profile is expected to exhibit a sequence of maxima and minima, the amplitudes of 
which should decay to zero when approaching the inner bulk-like portion of the film
\cite{mg,curro,pkryd,ykhs,freed,wy,wspcg,kr,theo3,bah,pakula,wang1,bithad,vaca1,kumar1,ron1,yet,yethall,vaca3}. 

Figure~\ref{moden} illustrates this behavior of $\rho_\mr{m}(z)$ at $T=\infty$ and $T=0.2$ for $D=30$ (see also
\cite{confit_bmb}). 
At $T=\infty$, the comparison of the low polymer, but high monomer concentration at the wall suggests that most of
the monomers at $z=1$ belong to different chains. This conclusion is true, as the calculation of the average number 
of monomers, which belong to the same chain and are in the same layer, shows. The number is about 3 at $T=\infty$,
but increases to 7 or 8 at $T=0.2$ (see also \cite{composit,psitf}). This is again evidence for the tendency of the 
chains to orient parallel to the wall under the influence of the model's energy function.

\section{Dynamic Properties of the Polymer Films}
\label{dynamic}
Previous work on the dynamics of polymer films focused 
on the behavior of mean-square displacements and related
quantities, such as the monomer mobility and the chain's diffusion
coefficient parallel to the walls \cite{compodyn,feature,mrs_98}.
Another important means to study dynamic properties are 
time-displaced correlation functions. The present section 
discusses two different kinds of these functions: the 
incoherent intermediate scattering function which probes 
density fluctuations, and the correlation functions of the Rouse 
modes which are sensitive to reorientations along the backbone 
of a chain.

\subsection{Incoherent Intermediate Scattering Function}
Let $\vec{r}_m(t)$ be the position vector to the $m$th monomer 
at time $t$. The incoherent scattering function can then be 
written as
\begin{equation}
\Phi_q^\mr{s}(t) = \frac{1}{M} \sum_{m=1}^M \Big \langle
\exp \Big(\mr{i}\vec{q} \cdot [\vec{r}_m(t) -\vec{r}_m(0)]
\Big)\Big\rangle \;,
\label{defphis}
\end{equation}
where $\vec{q}=(q_x,q_y,0)$ denotes a wave vector corresponding
to the $x,y$-plane of the simple cubic lattice and $\langle 
\bullet \rangle$ stands for both the thermodynamic average and
the average over all wave vectors with the same modulus $q=
|\vec{q}|$. Physically, the incoherent scattering function
measures displacements of a monomer in time. The dominant 
contribution to the decay of $\Phi_q^\mr{s}(t)$ comes from 
motions of the order $2\pi/q$.

\vspace{4mm}

\noindent
{\bf Qualitative Aspects of the Decay.}
An analysis at high temperature (i.e., at $T=\infty$) 
pointed out that the influence of finite film 
thickness on the decay of $\Phi_q^\mr{s}(t)$
diminishes with decreasing $q$-value, i.e., when density 
fluctuations on larger and larger length scales are probed
\cite{confit_bmb}. Thus, the present study focuses on
the peak of the static structure factor, $q=2.94$ \cite{cool2}, 
which corresponds to the scale of a bond in real space.

Figure~\ref{fig:phis} shows the time-dependence of 
$\Phi_q^\mr{s}(t)$ at this $q$-value for various temperatures 
and two different film thicknesses, $D=6$ ($\approx 1.5 R_\mr{g}$)
and $D=30$ ($\approx 7.5 R_\mr{g}$). The figure clearly 
illustrates the slowing down of structural relaxation as 
the temperature decreases from the normal liquid ($T=0.35$) 
to the supercooled state ($T=0.2$) of the melt. The term 
``supercooled state'' refers to temperatures close to, but still 
above the critical temperature of mode-coupling theory (MCT) 
\cite{goetze4,goetze1,goetze2,goetze3} for the bulk, which was 
estimated as $T_\mr{c} \approx 0.15$ in \cite{cool3,exmct}%
\footnote{Contrary to the present simulation data, the 
bulk configurations, from which the estimate $T_\mr{c} \approx 
0.15$ was derived, were not fully equilibrated. It is 
possible that a complete removal of these residual non-equilibrium 
effects shifts the result for $T_\mr{c}$ (to slightly higher
temperatures).}.

Two observations can be made from Fig.~\ref{fig:phis}: First, 
the thin film always relaxes faster than the thick film. This difference 
increases with decreasing temperature. Second, there is 
hardly any influence on the shape of the decay for both film 
thicknesses as long as $T > 0.2$. If $T=0.2$, a shoulder begins to emerge at
intermediate times for $D=30$. A similar feature is not visible
for $D=6$. This shoulder can be interpreted as the onset of the MCT-$\beta$ 
process whose characteristic signature is a two-step relaxation of 
$\Phi_q^\mr{s}(t)$ \cite{goetze4,goetze1,goetze2,goetze3}. Physically, the 
$\beta$-process corresponds to the
relaxation of particles in ``cages'' formed by their nearest neighbors.
As temperature approaches $T_\mr{c}$, a particle is trapped for some
time in its local enviroment (= ``cage'') before it can escape and
diffuse to an adjacent cage \cite{goetze4,goetze1,goetze2,goetze3}. 
The approximate theoretical treatment of this picture compares fairly well 
with simulation data for the bulk \cite{acs98,kb_rev93,cool3,exmct}. 
The present results suggest that this relaxation behavior might also 
develop in a thin film geometry at lower temperatures than studied.

\vspace{4mm}

\noindent
{\bf Scaling Behavior of the Incoherent Scattering Function.}
In order to illustrate these qualitative properties from a different
point of view we determined the relaxation time $\tau_q^\mr{s}$ from
the correlators by posing $\Phi_q^\mr{s}(\tau_q^\mr{s})=0.629$
and plotted $\Phi_q^\mr{s}(t)$ versus $t/\tau_q^\mr{s}$. Such a 
representation shifts the curves on top of each other and thus allows 
a better comparison of the influence of temperature and thickness on the 
shape of the decay. As expected from the qualitative discussion above, 
the data for both film thicknesses collapse at high temperature 
(exemplified by $T=0.35$ in Fig.~\ref{fig:phis_scale}). This indicates
that the shape is unaffected by the confinement. The main influence is 
a change of the relaxation time.

However, such a collapse is only possible for the final decay, i.e., the
late time $\alpha$-process, if the temperature belongs to the 
supercooled regime ($T \lesssim 0.23$). The shoulder is now clearly 
visible for $T=0.2$ and $D=30$ at intermediate times. Exactly in this
intermediate time window the simulation results for $D=6$ markedly 
deviate from $D=30$, whereas they remain (very) close to $D=30$ 
if $T>0.2$. The data for $D=6$ and $T=0.2$ do not exhibit the
shoulder. They rather resemble simulation results of $D=30$
obtained at a higher temperature than $T=0.2$. In other words, the
thin film seems to behave like the thick film at a larger temperature.
Therefore, the present analysis suggests that a possible influence of geometric
confinement on the dynamics of the polymer films could be
a shift of the temperature scale to higher values.
Evidence in favor of this interpretation is provided by
simulation results for a bead-spring model of a glassy
polymer film \cite{confit_vbb} and by the temperature
dependence of the relaxation time $\tau^\mr{s}_q$.

\vspace{4mm}

\noindent
{\bf Temperature and Thickness Dependence of the Relaxation
Time.}
Figure~\ref{fig:tauqs} shows the temperature dependence of 
the relaxation time $\tau_q^\mr{s}$ for the both film 
thicknesses, $D=6$ and $D=30$, and compares it with that of
bulk. One can see that the curves gradually splay out with 
decreasing temperature. The films always relax faster than 
the bulk, and the relaxation time of the film is the 
smaller, the smaller its thickness. If one extrapolates 
this trend to low temperature by a Vogel-Fulcher equation
\cite{jaeckle},
\begin{equation}
\tau_q^\mr{s} = A^\mr{s}_q \exp \left(\frac{E^\mr{s}_q}
{T-T_0}\right ) \;,
\label{eq:vft}
\end{equation}
one obtains a Vogel-Fulcher temperature $T_0$ which
decreases with decreasing film thickness \cite{confit_bmb}.
Therefore, density fluctuations on the length scale of bond
suggest a reduction of the glass transition temperature 
with decreasing film thickness.

\subsection{Rouse Modes}
Possible quantities to study reorientations are 
correlation functions of the Rouse modes \cite{doi}.
The Rouse modes are the cosine transforms of the position 
vectors, $\vec{r}_n$, to the mo\-no\-mers. For the discrete 
polymer model under consideration they can be written as 
\cite{verdier}
\begin{equation}
\vec{X}_p(t)=\frac{1}{N}\sum_{n=1}^{N} \vec{r}_n(t)
\cos\left (\frac{(n-1/2)p\pi}{N}\right ) \; , \quad 
p=0,\ldots,N-1 \; .
\label{defx}
\end{equation}
One can think of the Rouse modes as standing waves 
along the backbone of the chain. The first mode has nodes 
only at the ends and is thus sensitive to reorientations
of the whole polymer. The second mode has an additional 
node in the middle. It roughly divides the chain into two 
halves and probes reorientations of these two segments. 
The higher Rouse modes further decompose the chain. Approximately,
one can consider the $p$th mode as a quantity which is sensitive 
to orientational dynamics of a chain segment with $N/p$ monomers. 
For the studied chain length $N=10$ the fifth Rouse mode thus
measures relaxation processes on the
same length scale of $\Phi_q^\mr{s}(t)$ at the maximum of
the static structure factor. Therefore, we concentrate
on this mode.

The auto-correlation function of the fifth mode is given by
\begin{equation}
\Phi_{55}(t)=\frac{\langle \vec{X}_5(t)\cdot \vec{X}_5(0)\rangle}
{\langle \vec{X}_5(0)^2\rangle} = 
\exp\left (-\frac{t}{\tau_5}\right)\; , 
\label{defcx}
\end{equation}
where $\tau_5$ is the relaxation time of the mode. 
The second equality of Eq.~(\ref{defcx}) is the prediction 
of the Rouse model which is only approximately satisfied for the 
present model \cite{kbmb,owbb} in the bulk (and other models as well 
\cite{bbpb}), if $p=1$. For higher mode indices deviations from a 
simple exponential behavior are observed \cite{kbmb,owbb}. So, we 
expect to find similar differences between the simulation data for
$\Phi_{55}(t)$ and $\exp(-t/\tau_5)$ for the polymer films studied
here.

Since the Rouse modes are vectors, one can distinguish
between a component parallel and perpendicular to the wall.
In the following the auto-correlation function parallel
to the wall shall be considered only. Why? Imagine a polymer
film with a thickness of one monomer. In this 
two-dimensional geometry no relaxation perpendicular, but 
only parallel to the walls can occur. In a film of 
finite thickness one may expect that the parallel component
still dominates the dynamics of the correlation
function (\ref{defcx}) provided the thickness does not
reach bulk-like dimensions. Therefore, we focus our 
attention on the relaxation of the parallel component of 
the fifth mode, $\Phi_{55,\|}(t)$, in the remainder of
this section.

\vspace{4mm}

\noindent {\bf Qualitative Properties of the Decay.}
Figure~\ref{fig:phi55} shows the time-dependence of 
$\Phi_{55,\|}(t)$  for the same temperatures and film 
thicknesses as studied before for the incoherent scattering 
function. Again, several observations can be made: The 
influence of film thickness on the shape of the decay is 
almost negligible. It is still weaker than for 
$\Phi_{q}^\mr{s}(t)$. Throughout the temperature range
studied the thick film ($D=30$) remains very close to 
the bulk, and $\Phi_{55,\|}(t)$ always decays in a single
step for both film thicknesses and for the bulk%
\footnote{This could be a property of the studied
model because a two-step decay of the Rouse modes was
found in a molecular-dynamics simulation of a bead-spring model
\cite{bbpb}.}. There is no indication of a two-step relaxation 
contrary to $\Phi_q^\mr{s} (t)$. The film thickness seems to 
shift the relaxation time only. 

This is also illustrated in Fig.~\ref{fig:ttsp_rouse}
which shows an attempt to scale all data onto a master curve
by plotting $\Phi_{55,\|}(t)$ versus $t/\tau_{5}$ in 
analogy to Fig.~\ref{fig:phis_scale}. As in the case of 
$\Phi_q^\mr{s}(t)$, the relaxation time was defined via
$\Phi_{55,\|}(\tau_{5})=0.629$. This scaling almost
perfectly collapses all data onto a common curve, the 
late time decay (i.e., $\Phi_{55,\|}(t)< 0.5$) being
eventually slightly different between the supercooled state
($T\lesssim 0.23$) and the normal liquid state ($T\gtrsim 0.35$). 
This indicates that the confinement effects visible in 
Fig.~\ref{fig:phi55} leave the shape of $\Phi_{55,\|}(t)$, but 
only affect the relaxation time.

\vspace{4mm}

\noindent {\bf Temperature and Thickness Dependence of the 
Relaxation Time.}
Figure~\ref{fig:taupT} shows the temperature and thickness
dependences of $\tau_{5}$. Contrary to $\tau_{q}^\mr{s}$, 
one finds that $\tau_{5}$ increases with decreasing film 
thickness. The thin film ($D=6$) always relaxes more slowly
than the thick film ($D=30$) which is in turn (a bit) 
slower than the bulk. A tentative explanation could perhaps
be as follows: The discussion of the static properties 
showed that chains in the film of thickness $D=6$ are 
oriented parallel to the walls. At high temperature they
crowd in the middle, whereas an oscillatory density
profile with a high concentration at the walls develops 
during supercooling. In both cases the decorrelation of 
$\vec{X}_{5,\|}$ has to take place by many parallel 
displacements of all monomers of a chain in an environment 
where the chain density is higher than in the thick film or
in the bulk. So, the relaxation could be slowed down for 
$D=6$ in comparison with $D=30$ which is more homogeneous 
and thus more bulk-like. 

\section{Summary}
\label{sum}
This paper reports simulation results for a simple model of
glassy polymer films. The model consists of short 
(non-entangled) monodisperse chains. The monomers of
the chains solely interact by excluded volume forces with
each other and with two completely smooth walls that
define the thin film geometry. The glassy behavior is
brought about by a competition between the internal energy
of a chain, which tries to make the chain expand at low
temperatures, and the dense arrangement of all monomers in 
the available volume. The temperatures investigated are
taken from the temperature interval above the bulk critical
temperature of mode-coupling theory. 

With this model we studied both static and dynamic features
of polymer films of various thicknesses. For the interpretation 
of the static results
it is important to realize that the instantaneous shape of 
a polymer resembles an ellipsoid. Close to a hard wall, the 
ellipsoid tends to orient parallel to it (see
Fig.~\ref{fig:sketch1}). If the film 
thickness is large enough ($D \gtrsim 5 R_\mr{g}$), there is
a bulk-like inner region with free orientation of the 
chains. But, as the film thickness decreases, this 
orientational freedom becomes more and more limited. If $D 
\lesssim 4 R_\mr{g}$, the perturbations of the structure 
induced by both walls interfere so that chains also align 
parallel to the walls in the inner portion 
of the film. With decreasing temperature this wall effect
becomes strongly enhanced as a consequence of the internal
energy of the chains which favors bond vectors that point
along the lattice axes and are thus parallel to the wall.
 
The deviations from the isotropic structure of the bulk
also affect the dynamic properties of the films. The
relaxation of vectors, such as the Rouse modes, is dominated
by reorientations parallel to the wall due to the 
alignment of the chains if the film thickness is small or if
the temperature is low. Under these conditions (small
$D$ and $T$) reorientations occur in an environment of
high chain density which might be the reason why the 
relaxation time of orientational correlation functions
increases with decreasing film thickness.

On the other hand, these orientations could also influence
local density fluctuations. Oriented chains should have a 
smaller number of contacts with other chains compared to 
the bulk because they are less intermingled. This might be
an explanation for the faster relaxation of $\phi_q^\mr{s}(t)$
in the thin film ($D=6$) already at high temperature and 
particularly at low temperature, where orientation effects
become more pronounced due to the impact of the model's
energy function (see also \cite{nrp} for a similar argument 
to interprete experimental results for freely-standing polymer 
films).

At low temperatures, however, another effect could also contribute. 
Theoretical developments \cite{goetze1,goetze3} as well as tests 
by experiments and simulations \cite{goetze4,goetze2,yip,kob_rev95,%
kob_rev99} have suggested
that a significant contribution to the slowing down of 
structural relaxation stems from the mutual blocking of 
particles which are close in space (particles in the
``cage''). This cage effect becomes important for the 
dynamics in the supercooled state and is determined by the 
liquid structure of the bulk, particularly by the first 
neighbor shell. Deviations from bulk behavior should be
observed if this structure is perturbed. The structure is perturbed
in the polymer films studied. The completely smooth walls 
of our model cut off the liquid structure and thereby remove 
part of the obstacles (i.e., other particles) which impede 
the displacement of a tagged particle (see Fig.~\ref{fig:sketch2}). 
The walls act like a lubricant with respect to the bulk, which 
enhances the mobility of the nearby monomers with respect to the bulk. 
These more mobile monomers can transfer part of this impetus towards 
the inner portion of the film so that the effect should propagate over 
a certain distance away from the wall before it is gradually damped out. 

This effect could complement the orientation-induced reduction of 
contacts for thin films in the supercooled state where the dynamics is 
very cooperative. Even if the chains densely populate the layer next to 
the wall at low temperatures for small film thickness and thus essentially 
form a more or less rugged surface in front of the wall, this surface is 
not frozen in, as assumed in the right panel of Fig.~\ref{fig:sketch2}, for
instance. On the one hand, this surface makes adjacent chains to orient 
parallel to it, thereby effectively reducing the number of contacts, and on 
the other hand, it should be lubricated by the smooth underlying wall and pass 
part of this enhanced mobility over to neighboring layers. Both effects could 
therefore reinforce one another.

However, there is still another effect which might eventually outweigh
this enhanced mobility. The discussion of the static properties showed
that the monomer density is larger at the wall than in the bulk and
decays towards the bulk density in an oscillatory fashion with 
increasing distance from the wall. If the external control parameters 
(temperature, pressure, film thickness) are such that the density 
oscillations become very pronounced, the film could be divided into 
highly populated layers separated by sharp depletion zones. This seems
to be the case in the above cited simulations of hard sphere mixtures
\cite{fl1995,YamamotoKim2000}. At low temperature the particle density 
at the wall is about a factor of 4 larger than the bulk value in these
studies. In the present simulation, it is only a factor of about 2. It is 
plausible that the mobility should decrease if the density surmounts a
certain threshold, which could in turn slow down the dynamics of the 
whole film. Therefore, even for a completely smooth wall the extent of
the structuration of the liquid close to it might lead to faster or
slower dynamics compared to the bulk. Qualitatively, this argument is 
similar to those presented in a discussion of the influence of wetting 
properties on diffusion in confined liquids 
\cite{BarratBocquet1999,Almerasetal2000}.

Finally, the preceding discussion always assumed that there is no
preferential attraction between the monomers and the walls. If such an
attraction, however, exists and if it is strong, one would expect a 
rather immobile interfacial layer of adsorbed particles to form. For small
film thicknesses this layer should considerably influence of the overall
film dynamics and could contribute to slowing down the structural relaxation 
of the film. In this case, the glass transition temperature should increase for
thin films with respect to the bulk. There is evidence for this behavior 
from simulations of both freely standing and supported polymer films 
\cite{TNdePablo} and from experiments of polymer films \cite{FNdePablo,kjcFaraday} 
(see also \cite{Kremeretal1999,asghk} for comparable results of molecular glass 
formers).

\section*{Acknowledgement}
We are indebted to S. Dasgupta, F. Eurich, T. Kreer, P. Maass, M. M\"uller, W. Paul and F. Varnik for helpful 
discussions on various aspects of this work and to the unknown referees for their valuable comments. This study 
would not have been possible without a generous grant of simulation time by the HLRZ J\"ulich, the RHRK
Kaiserslautern and the computer center at the University of Mainz. Financial support by the ESF Programme on 
``Experimental and Theoretical Investigation of Complex Polymer Structures'' (SUPERNET) 
is gratefully acknowledged.

%
%
\newpage
\begin{figure}
\caption[]{Sketch of the simulation geometry (left panel) 
and of the model (right panel). The simulation box is 
confined by two hard walls in the $z$-direction, which are 
a distance $D$ apart ($D=6,\ldots,60 \approx (1.5,\ldots,15)
R_\mr{g}$; $R_\mr{g}$: bulk radius of gyration). In the 
$x$- and $y$-directions periodic boundary conditions are 
used (exemplified by the bond leaving the bottom and 
reentering at the top). The linear dimension in these 
directions is $L=60$. The right panel shows a possible 
configuration of two different chains. All bonds have 
energy $\epsilon$ ($\epsilon/k_\mr{B}=1$: this defines the 
temperature scale) except the bond $(3,0,0)$ which is in 
the ground state (two-level system). This vector
blocks four lattice sites (marked by $\circ$)
due to the excluded volume interaction. This interaction
also forbids the jump in direction of the arrow.}
\label{coarse+hard}
\end{figure} 
\begin{figure}
\caption[]{Asphericity $\langle \Delta_0(z_\mr{cm}) \rangle$ of a chain versus the distance, $z_\mr{cm}$, of the
chain's center of mass from the (left) wall at $T=\infty$ (main figure). $\langle \Delta_0(z_\mr{cm}) \rangle$ is 
defined by Eq.~(\ref{defDelta}). Since the profile of $\langle \Delta_0(z_\mr{cm}) \rangle$ is symmetric around 
the middle of the film, only one half is shown. Various film thicknesses $D$, ranging from $D=6 \approx 1.5R_\mr{g}$ 
to $D=60 \approx 15 R_\mr{g}$, are compared ($R_\mr{g}(T=\infty) \simeq 3.68 =$ bulk radius of gyration; vertical 
dashed line). The result for the bulk, $\langle \Delta_0 \rangle = 
0.411$, is also depicted (horizontal filled circles $\bullet$). This value is close to that expected for a random walk 
$\langle \Delta_0 \rangle = 0.396$ \cite{cag}. The inset illustrates the temperature dependence of 
$\langle \Delta_0(z_\mr{cm}) \rangle$ for the thinnest film $D=6$. Here, the profile over the whole film is 
shown. The temperature varies from the high-$T$, liquid
state ($T=\infty$) to the supercooled state of the melt ($T=0.18$). Lines are guides to the eye only. All lengths 
are measured in units of the lattice constant, and temperature is measured in units of $\epsilon/k_\mr{B}$ (see 
Sect.~\ref{model} and Fig.~\ref{coarse+hard}).}
\label{aspher.allT.allD}
\end{figure}
\begin{figure}
\caption[]{Prolateness $\langle S_0(z_\mr{cm}) \rangle$ of a chain versus the distance, $z_\mr{cm}$, of the
chain's center of mass from the (left) wall at $T=\infty$ (main figure). $\langle S_0(z_\mr{cm}) \rangle$ is 
defined by Eq.~(\ref{defS}). Since the profile of $\langle S_0(z_\mr{cm}) \rangle$ is symmetric around
the middle of the film, only one half is shown.
Various film thicknesses $D$, ranging from $D=6 \approx 1.5R_\mr{g}$ to $D=60 \approx 
15 R_\mr{g}$, are compared ($R_\mr{g}(T=\infty) \simeq 3.68 =$ bulk radius of gyration; vertical dashed line). 
The result for the bulk, $\langle S_0 \rangle = 
0.499$, is also depicted (horizontal filled circles $\bullet$). This value is close to that expected for a random walk 
$\langle S_0 \rangle = 0.481$ \cite{cag}. The inset illustrates the temperature dependence of 
$\langle S_0(z_\mr{cm}) \rangle$ for the thinnest film $D=6$.  Here, the profile over the whole film is
shown.  The temperature varies from the high-$T$, liquid
state ($T=\infty$) to the supercooled state of the melt ($T=0.18$). Lines are guides to the eye only. All lengths 
are measured in units of the lattice constant, and temperature is measured in units of $\epsilon/k_\mr{B}$ (see 
Sect.~\ref{model} and Fig.~\ref{coarse+hard}).}
\label{prolat.allT.allD}
\end{figure}
\begin{figure}
\caption[]{Profile of the radius of gyration, $R_\mr{g}^2(z_\mr{cm})$, measured parallel
and perpendicular to the wall, at $T=\infty$ (main figure). $z_\mr{cm}$ denotes
the distance of the chains' center of mass from the left wall. Since the profiles are
symmetric around the middle of the film, the left half is only shown. Three different
thicknesses are presented: $D=6$ ($\approx 1.5R_\mr{g}$), $D=12$, and $D=30$
($\approx 7.5R_\mr{g}$). The bulk radius of gyration, $R_\mr{g}$ ($\approx 3.68$), is indicated 
as a vertical dashed line. All lengths are measured in units of the lattice constant. The inset
shows the counterpart of the simulation to the experimental results of \cite{kmkss,jkhbr} (see
Figs.~3 of \cite{kmkss,jkhbr}). Here, $R_{\mr{g},\|}(\mbox{film})$ denotes the average of the parallel
component of $R_\mr{g}(z_\mr{cm})$ over the whole film and $R_{\mr{g},\|}=2R_{\mr{g}}/3$ is 
the corresponding bulk value. The ratio $R_{\mr{g},\|}(\mbox{film})/R_{\mr{g},\|}$, plotted
versus $D/R_\mr{g}$, indicates that the average behavior of the whole film becomes bulk-like if
$D \gtrsim 5 R_\mr{g}$, as also observed in \cite{kmkss}. The dotted horizontal line shows 
the statistical uncertainty of $R_{\mr{g},\|}$ (horizontal dashed line). However, the overall expansion with
decreasing $D/R_\mr{g}$ is weaker in the simulation than in \cite{kmkss}.
}
\label{rgden}
\end{figure}
\begin{figure}
\caption[]{The main figure shows the variation of the chain-density profile, $\rho_\mr{c}(z_\mr{cm})$, with the
distance, $z_\mr{cm}$, of the chain's center of mass from the (left) wall in the high-temperature liquid state
of the films ($T=\infty$, i.e., only excluded volume interactions are effective). 
Since the profile is symmetric around the middle of the film, only one half is shown. Different film thicknesses,
ranging from $D=6$ ($\approx 1.5R_\mr{g}$) to $D=60$ ($\approx 15 R_\mr{g}$), are compared ($R_\mr{g}(T=\infty) 
\simeq 3.68 =$ 
bulk radius of gyration; vertical dashed line). The small horizontal circles ($\bullet$) indicate the bulk
value $\rho_{\mr{c},\mr{bulk}}=6.\bar{6}\times 10^{-3}$ ($=K/L^2 D$; see Sect.~\ref{model}). 
The inset illustrates the temperature dependence
of $\rho_\mr{c}(z_\mr{cm})$ for $D=6$. Here, the profile over the whole film is shown.
Lines are guides to the eye only. The temperature varies from the high-$T$, 
liquid state ($T=\infty$) to the supercooled state of the melt ($T=0.18$). All lengths are measured in units
of the lattice constant, and temperature is measured in units of $\epsilon/k_\mr{B}$ (see Sect.~\ref{model}
and Fig.~\ref{coarse+hard}).}
\label{chaindensity.allT.allD}
\end{figure}
\begin{figure}
\caption[]{Temperature dependence of the monomer density profile $\rho_\mr{m}(z)$ for $D=30 \approx 7.5 R_\mr{g}$. 
$z$ denotes the distance of the (lower left corner of the) monomer from the (left) wall.
The bulk radius of gyration $R_\mr{g}$ increases from $R_\mr{g}\simeq 3.68$ at $T=\infty$ to $R_\mr{g}\simeq 4.64$ 
at $T=0.2$ in the studied temperature interval. $T=\infty$ and $T=0.2$ are characteristic temperatures of the 
high-temperature, liquid state and the supercooled state of the melt, respectively. The horizontal circles 
($\bullet$) indicate the bulk value $\rho_{\mr{m},\mr{bulk}}=6.\bar{6}\times 10^{-2}$ ($=NK/L^2 D$; see 
Sect.~\ref{model}). Lines are guides to the eye only. All lengths are measured in units of the lattice constant, 
and temperature is measured in units of $\epsilon/k_\mr{B}$ (see Sect.~\ref{model} and Fig.~\ref{coarse+hard}).}
\label{moden}
\end{figure}
\begin{figure}
\caption[]{Time dependence of the incoherent scattering
function $\Phi_q^\mr{s}(t)$ for various temperatures
ranging from the normal liquid state ($T=0.35$) to the
supercooled state ($T=0.2$) of the melt. $\Phi_q^\mr{s}(t)$
is shown at the maximum of the static structure factor 
(i.e., at $q=2.94$ corresponding to a distance of 2 (lattice 
constants) in real space $\approx$ length scale of a
bond). Two film thicknesses, $D=6$ ($\approx 1.5 R_\mr{g}$;
depicted by symbols) and $D=30$ ($\approx 7.5 R_\mr{g}$;
depicted by lines) are compared. 
The correspondence between temperature and line type is 
indicated in the figure. For the symbols it is: 
$\circ$ ($T=0.35$), 
$\ast$ ($T=0.26$),
$\bigtriangleup$ ($T=0.23$),
$\Box$ ($T=0.2$).
Time is measured in 
Monte Carlo steps [MCS], i.e., the time needed to give
all monomers the chance to move once.
}
\label{fig:phis}
\end{figure}
\begin{figure}
\caption[]{Incoherent intermediate scattering function
$\Phi_q^\mr{s}(t)$ versus scaled time $t/\tau_q^\mr{s}$.
The scaling time is defined by the condition:
$\Phi_q^\mr{s}(\tau_q^\mr{s})=0.629$. The $q$-value 
corresponds to the maximum of the static structure factor
($q=2.94$), as in Fig.~\ref{fig:phis}. Three different
temperatures are shown ($T=0.35 \;\hat{=}$ normal liquid
state of the melt; $T=0.23,0.2\; \hat{=}$ supercooled state).
The symbols represent the data of the thin film $D=6$
($\approx 1.5 R_\mr{g}$) and the lines those of the thick 
film $D=30$ ($\approx 7.5 R_\mr{g}$). 
The correspondence between temperature and line type is indicated 
in the figure. For the symbols it is: $\circ$ ($T=0.35$), 
$\ast$ ($T=0.23$), $\Box$ ($T=0.2$).
A simple exponential, $\exp(-0.46 t/\tau_q^\mr{s})$, is also shown 
to illustrate
that the decay of $\Phi_q^\mr{s}(t)$ is always stretched.
The factor $a=0.46$ in the exponent results from the 
requirement $\Phi_q^\mr{s}(\tau_q^\mr{s})=\exp(-a)=0.629$.}
\label{fig:phis_scale}
\end{figure}
\begin{figure}
\caption[]{Temperature dependence of the relaxation times of $\Phi_q^\mr{s}(t)$
for two film thicknesses, $D=6$ ($\approx 1.5 R_\mr{g}$) and $D=30$ ($\approx 7.5 
R_\mr{g}$), and the
bulk (= unconfined system of linear dimension $L=60$). The relaxation time is
defined by $\Phi_q^\mr{s}(\tau_q^\mr{s})=0.629$ (see Fig.~\ref{fig:phis_scale}).
The lines represent fits to the Vogel-Fulcher equation [see Eq.~(\ref{eq:vft})]. 
The prefactor $A_q^\mr{s}$ of Eq.~(\ref{eq:vft}) should be proportional to
$\tau_q^\mr{s}(T=\infty)$. Since the variation of $\tau_q^\mr{s}(T=\infty)$ with 
$D$ is weak compared to that with $T$, it is possible to fit all curves with
a constant prefactor $A_q^\mr{s}=1.03 \pm 0.03$. Furthermore, the activation energy 
can also be taken as constant: $E_q^\mr{s}=0.855 \pm 0.005$. Only $T_0$ depends on $D$:
$T_0=0.119$ ($D=6$), $T_0=0.128$ ($D=30$), $T_0=0.136$ (bulk) (error bars for all $D$:
$T_0 + 0.001 /\!\! - 0.002$). The bulk results are compatible with earlier findings \cite{owbb}.}
\label{fig:tauqs}
\end{figure}
\begin{figure}
\caption[]{Time dependence of the auto-correlation 
function of the fifth Rouse mode, $\Phi_{55,\|}(t)$,
measured parallel to the wall. Different temperatures
are shown, which range from the normal liquid state 
($T=0.35$) to the supercooled state ($T=0.2$) of the melt. 
The correspondence between temperature and line type is 
indicated in the figure. For the symbols it is: 
$\circ$ ($T=0.35$), 
$\ast$ ($T=0.26$),
$\bigtriangleup$ ($T=0.23$),
$\Box$ ($T=0.2$).
Two film thicknesses, $D=6$ ($\approx 1.5 R_\mr{g}$;
depicted by symbols) and $D=30$ ($\approx 7.5 R_\mr{g}$;
depicted by lines) are compared with each other and with
the bulk (indicated by $\times$ for $T=0.35$ and $T=0.2$). 
Time is measured in Monte Carlo steps [MCS].}
\label{fig:phi55}
\end{figure}
\begin{figure}
\caption[]{Same data for $\Phi_{55,\|}(t)$ as in 
Fig.~\ref{fig:phi55}, but plotted versus scaled time 
$t/\tau_5$. The scaling time is defined by the condition:
$\Phi_{55,\|}(\tau_5)=0.629$. 
The symbols represent the data of the thin film $D=6$
($\approx 1.5 R_\mr{g}$) and the lines those of the thick 
film $D=30$ ($\approx 7.5 R_\mr{g}$). 
The correspondence between temperature and line type is indicated 
in the figure. For the symbols it is: $\circ$ ($T=0.35$), 
$\bigtriangleup$ ($T=0.23$), $\Box$ ($T=0.2$).
Furthermore, the results for the bulk are also included for 
$T=0.2$ ($\times$) and $T=0.35$ ($\bullet$).
A simple exponential, $\exp(-0.46 t/\tau_5)$, is also shown 
to illustrate that the decay of $\Phi_{55,\|}(t)$ is always stretched.
The factor $a=0.46$ in the exponent results from the 
requirement $\Phi_{55,\|}(\tau_5)=\exp(-a)=0.629$.}
\label{fig:ttsp_rouse}
\end{figure}
\begin{figure}
\caption[]{Temperature dependence of the relaxation times, $\tau_5(T)$, of 
the fifth Rouse mode $\Phi_{55,\|}(t)$. The relaxation time is
defined by $\Phi_{55,\|}(\tau_5)=0.629$. The ordinate is divided by the 
infinite temperature result $\tau_5(\infty)$ to illustrate the influence
of film thickness with decreasing temperature. Two film thicknesses, $D=6$ 
($\approx 1.5R_\mr{g}$) and $D=30$ ($\approx 7.5 R_\mr{g}$), are compared with the
bulk (= unconfined system of linear dimension $L=60$).} 
\label{fig:taupT}
\end{figure}
\begin{figure}
\caption[]{Schematic sketch of the influence of 
film thickness and temperature on static properties.
A good approximation of the instantaneous structure of a 
chain is an ellipsoid. The chains are therefore represented
by ellipsoids in the figure. Adjacent to a wall the 
two largest principal axes of the ellipsoid are aligned 
parallel to the wall, whereas the smallest 
axis points away from it in perpendicular direction.
Furtherfore, the chain is distorted (i.e., flattened). 
Both effects are entropically unfavorable so that the
number of chains close to the wall is small at high
temperature. If the film is sufficiently large, there is 
a bulk-like inner region, where the chains can freely
reorient (indicated by dashed ellipsoids in the left
sketch). With decreasing film thickness the bulk-like
region shrinks more and more, leading finally to parallel 
oriented chains also in the middle of the film if the
thickness is $D \lesssim 4 R_\mr{g}$. At high temperature
the walls are still avoided in such a thin film so that a 
high concentration
of chains occurs in the middle (indicated by the three
ellipsoids in the right sketch instead of two in the left).
Furthermore, the chains stiffen at low temperatures. This 
additionally favors orientation parallel to the walls, leading 
to a sequence of highly and weakly populated lattice layer for 
our model.}
\label{fig:sketch1}
\end{figure}
\begin{figure}
\caption[]{Illustration of a possible influence 
of confinement on the dynamics of a supercooled liquid.
Two types of confinement are compared with the bulk:
completely smooth walls and rough walls which are perfectly 
adapted to the liquid structure. In both cases, we assume that
the walls are neutral. There is no strong attraction which leads
to adsorption of the particles. Let us consider
the bulk first and imagine that the particle $\bigcirc$ moves
from its dashed position in direction of the large arrow. It 
opens space which its neighbors try to occupy. The dark-shaded 
particle has to compete with all other neighbors which are 
simultaneously moving in the space formerly occupied by
$\bigcirc$. This exerts a friction on the dark particle, which 
is partly absent if it is situated next to a completely smooth wall 
(panel in the middle). On the other hand, the friction should
be strongly enhanced if the particle is next to a rough wall
exhibiting cavities, into which it perfectly fits (right panel).
This contraint does not relax away
contrary to the bulk. So, one can expect the mobility
of dark particle to decrease with respect to the bulk for 
such a rough wall (see \cite{skb,confit_skb}, for instance), but to 
increase for a completely smooth wall which acts like a lubricant 
compared to the bulk.}
\label{fig:sketch2}
\end{figure}
%
%
\newpage
\setcounter{figure}{0}
\begin{figure}
\epsfysize=80mm
\hspace*{-15mm}\epsffile{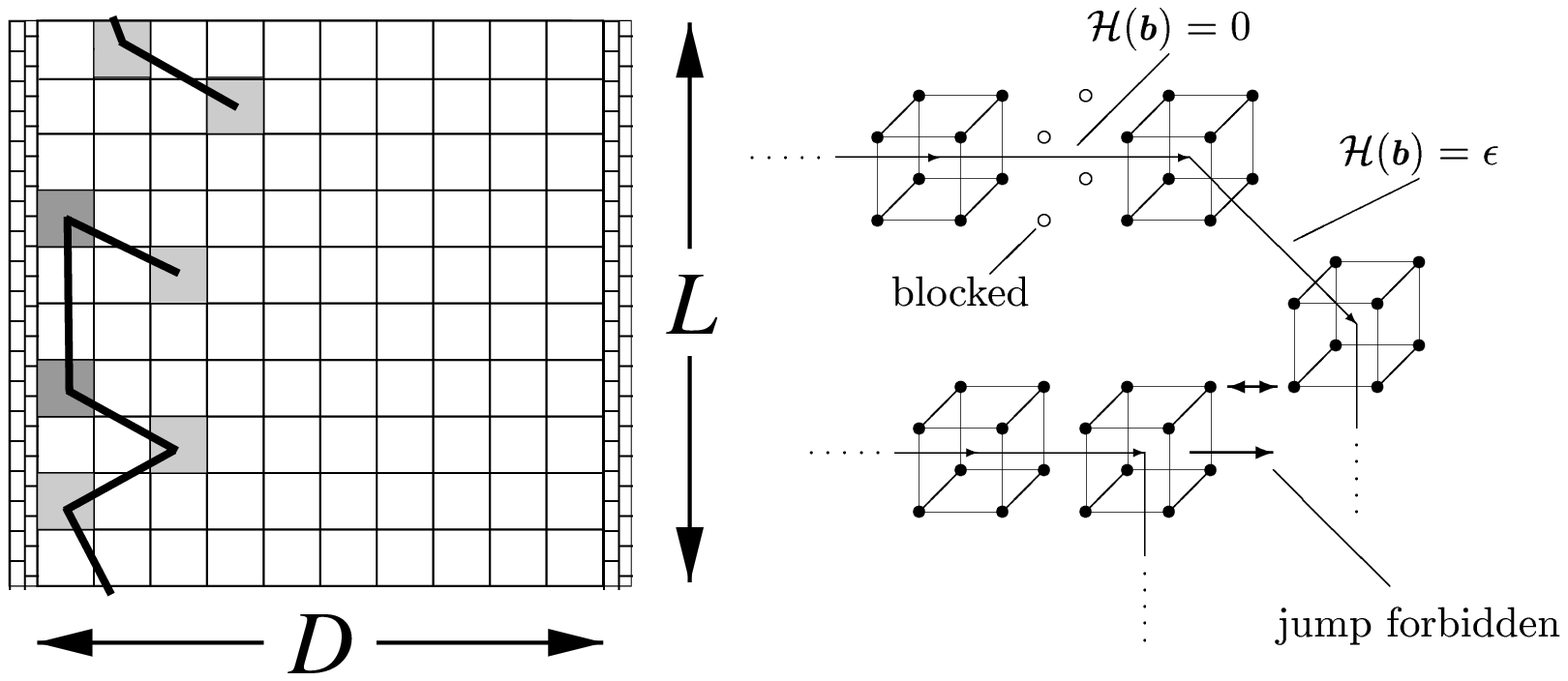}
\vspace{7mm}
\caption[]{}
\end{figure} 
\newpage
\begin{figure}
\epsfysize=140mm
\hspace*{-20mm}\epsffile{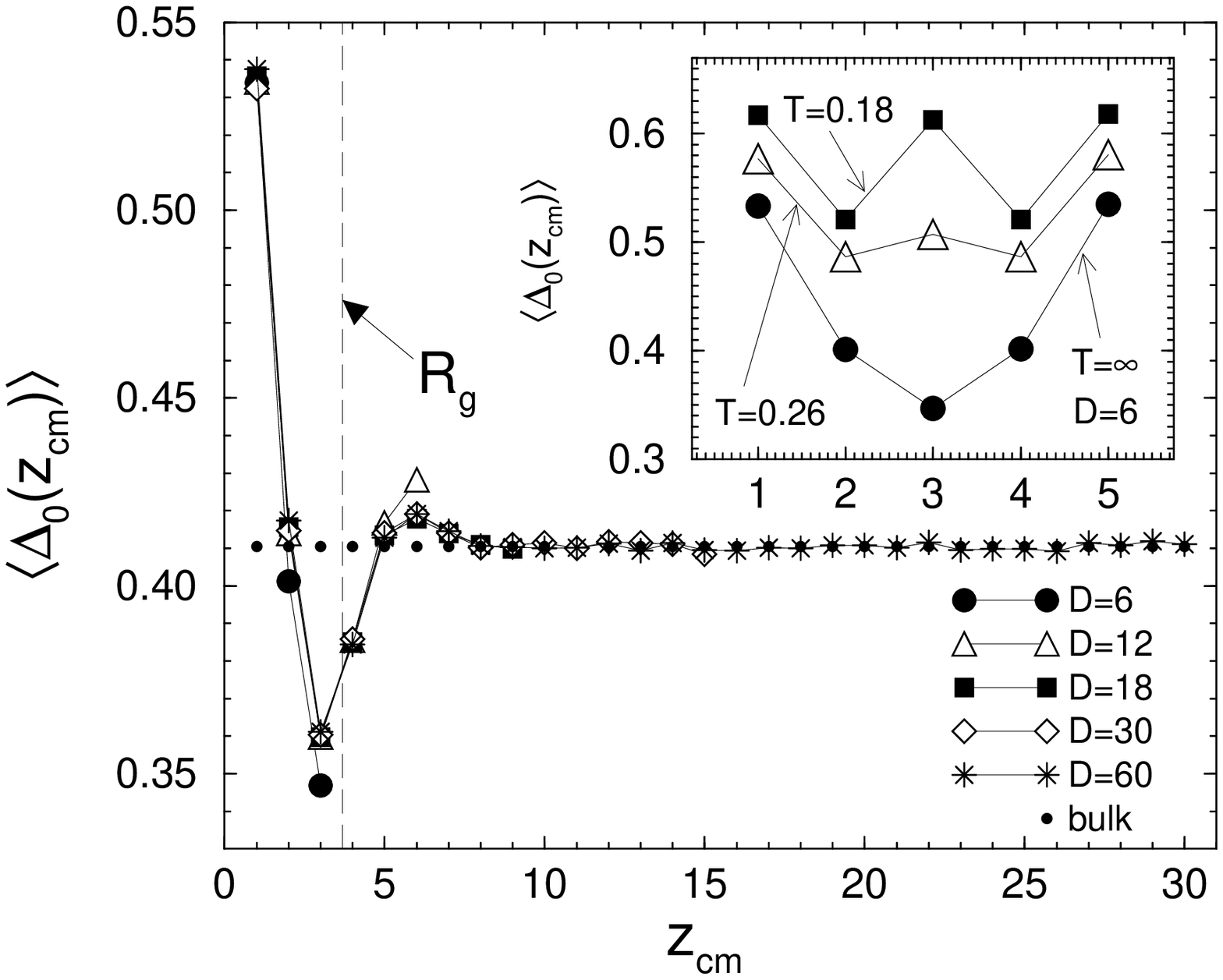}
\vspace{4mm}
\caption[]{}
\end{figure}
\begin{figure}
\epsfysize=140mm
\hspace*{-20mm}\epsffile{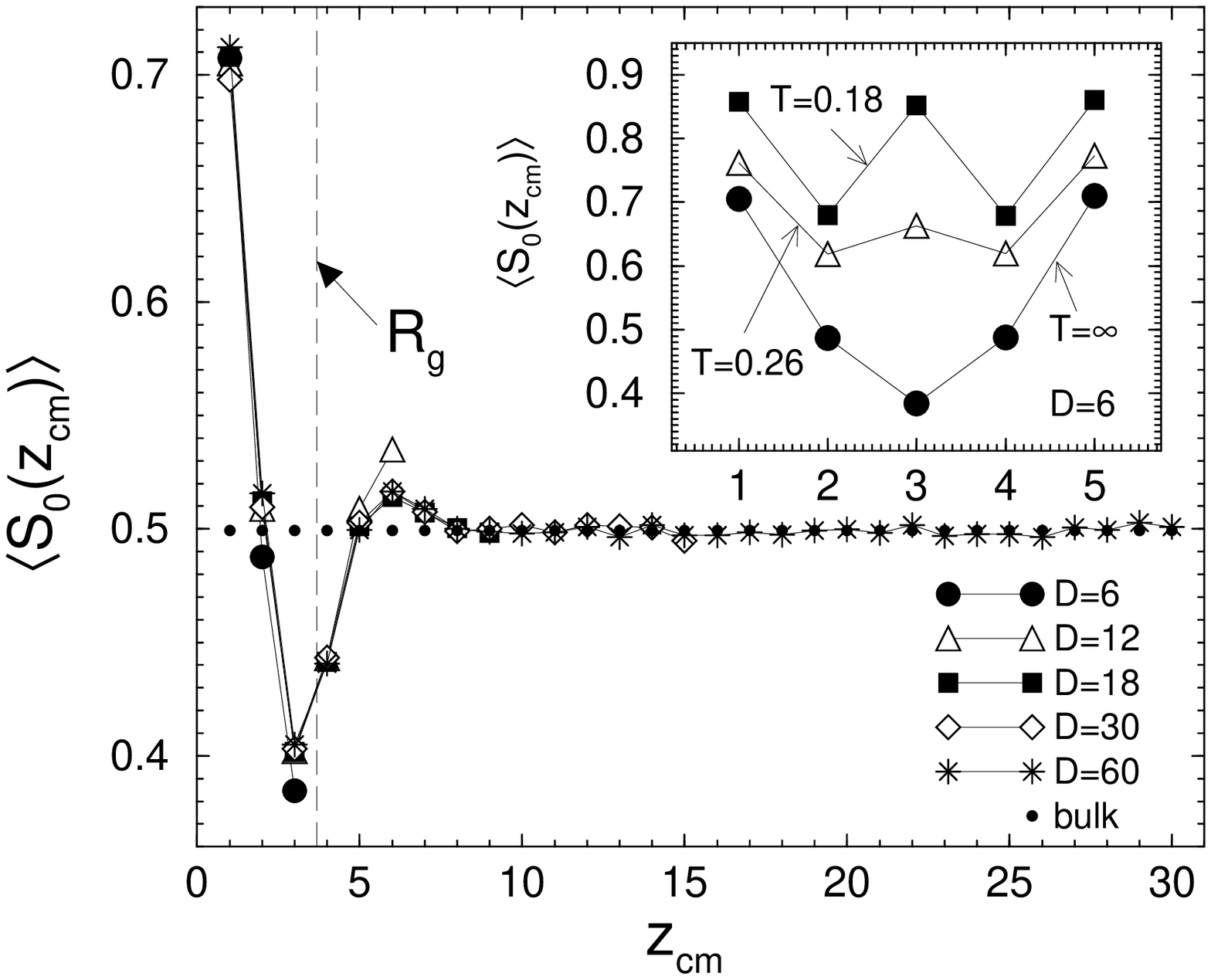}
\vspace{4mm}
\caption[]{}
\end{figure}
\begin{figure}
\epsfysize=140mm
\hspace*{-20mm}\epsffile{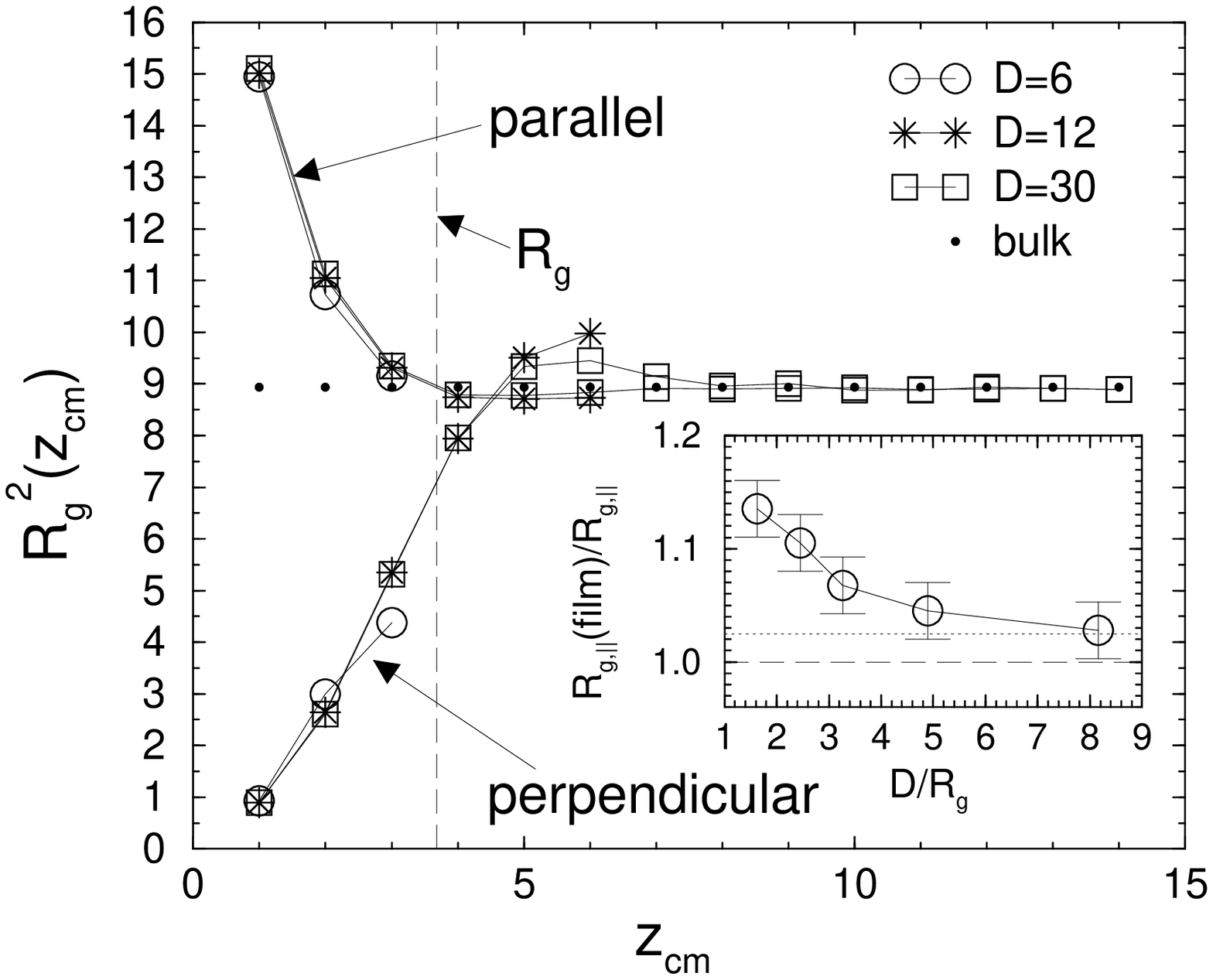}
\vspace{4mm}
\caption[]{}
\end{figure}
\begin{figure}
\epsfysize=140mm
\hspace*{-20mm}\epsffile{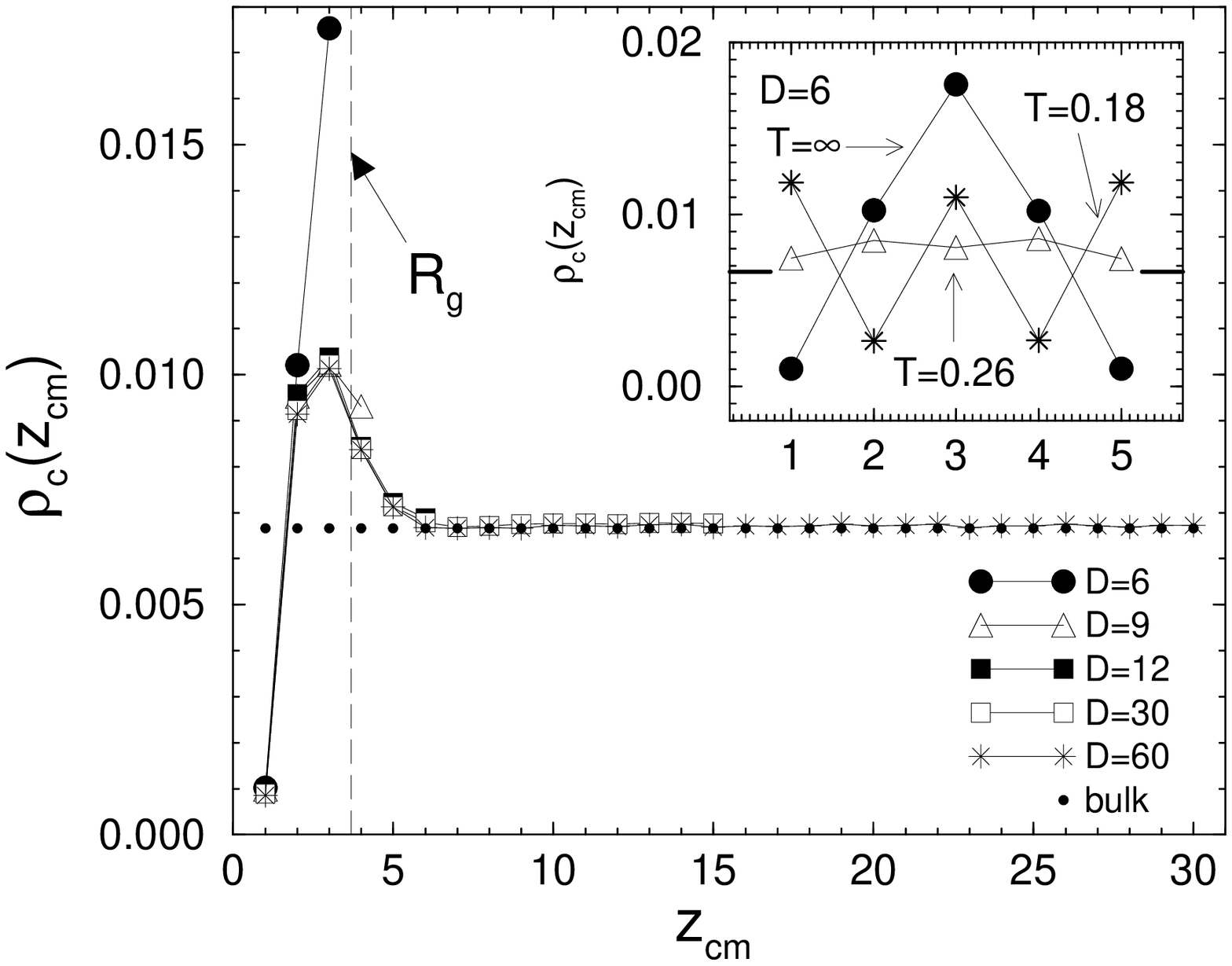}
\vspace{4mm}
\caption[]{}
\end{figure}
\begin{figure}
\epsfysize=140mm
\hspace*{-20mm}\epsffile{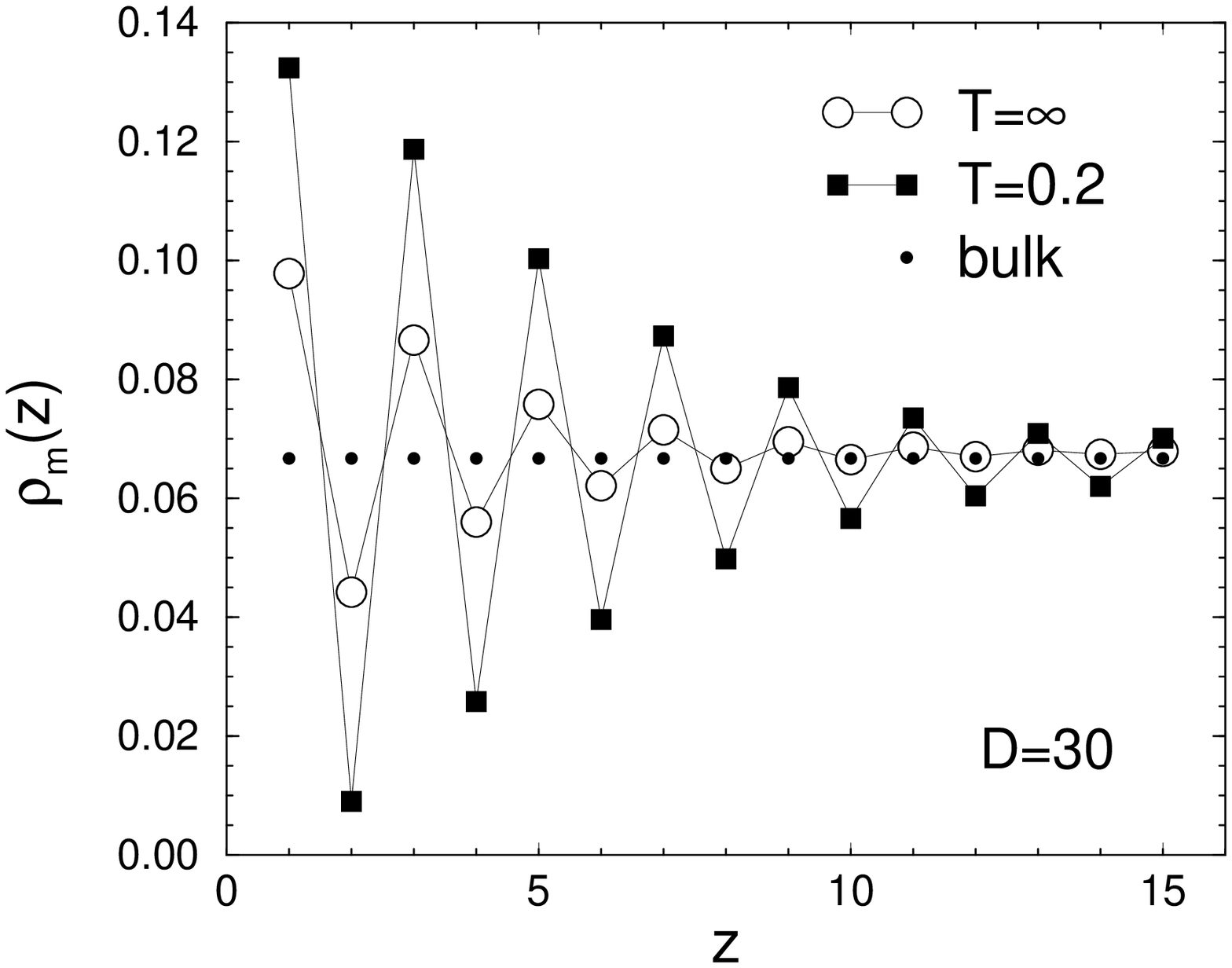}
\vspace{4mm}
\caption[]{}
\end{figure}
\begin{figure}
\epsfysize=140mm
\hspace*{-20mm}\epsffile{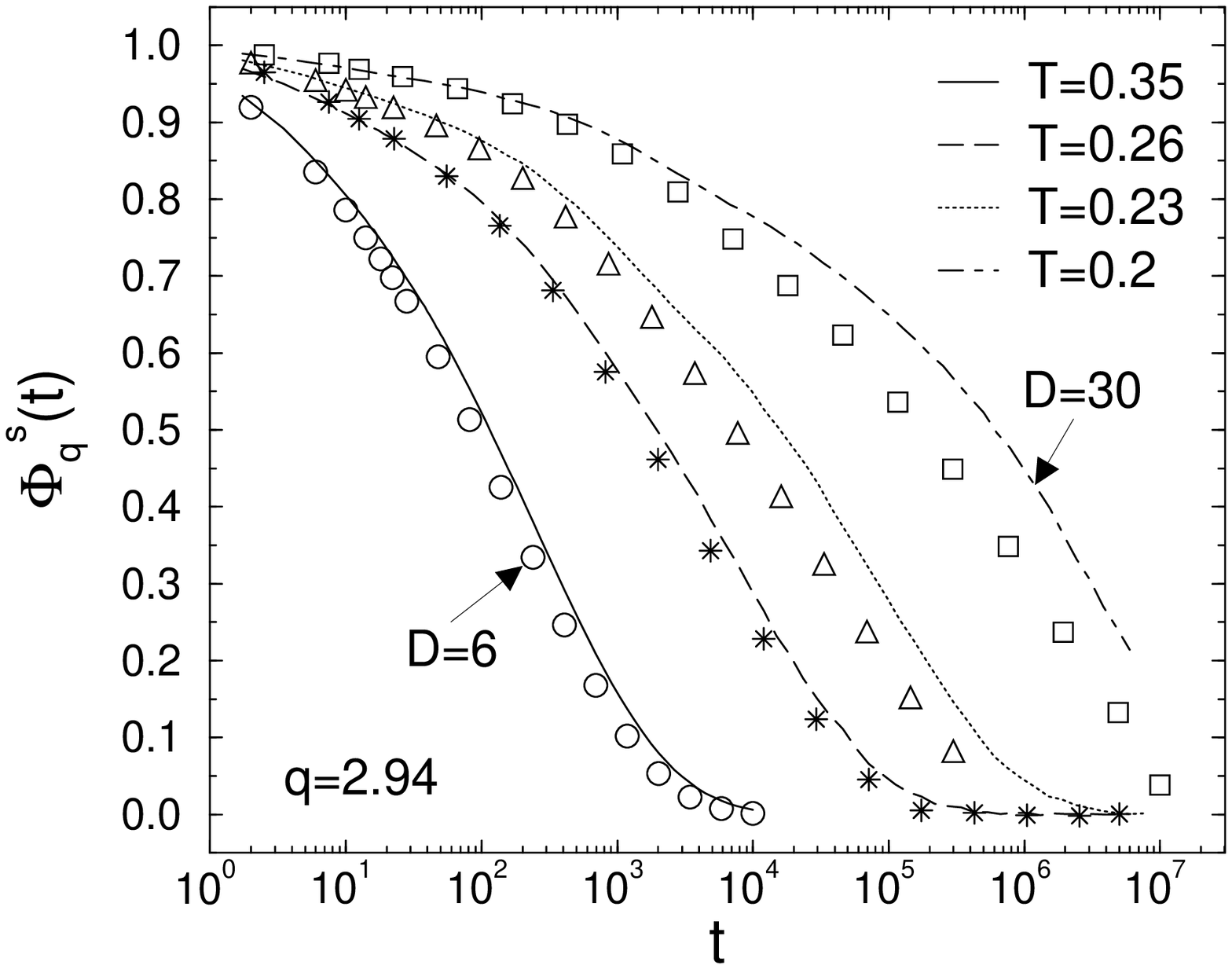}
\vspace{4mm}
\caption[]{}
\end{figure}
\begin{figure}
\epsfysize=140mm
\hspace*{-20mm}\epsffile{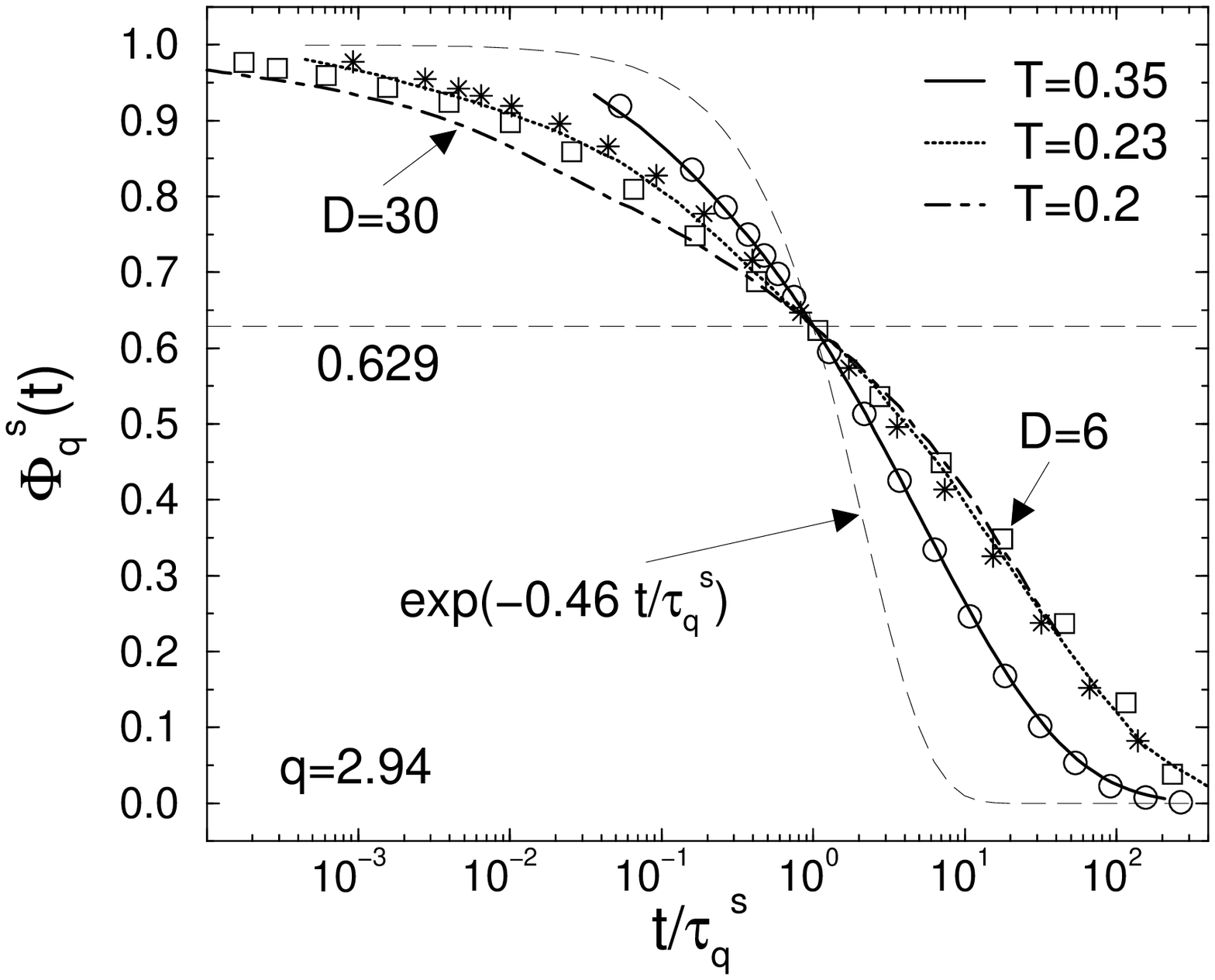}
\vspace{4mm}
\caption[]{}
\end{figure}
\begin{figure}
\epsfysize=140mm
\hspace*{-20mm}\epsffile{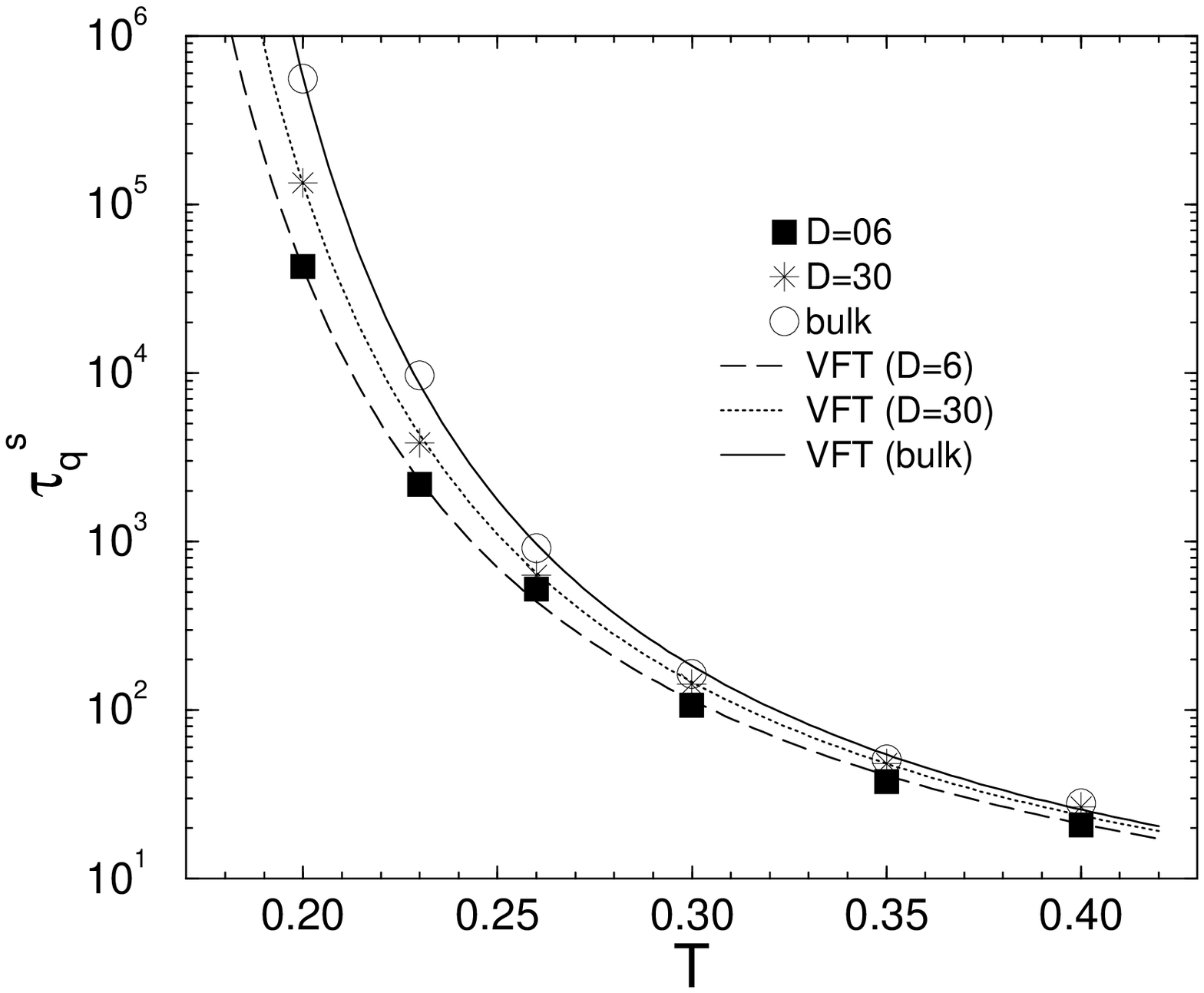}
\vspace{4mm}
\caption[]{}
\end{figure}
\begin{figure}
\epsfysize=140mm
\hspace*{-20mm}\epsffile{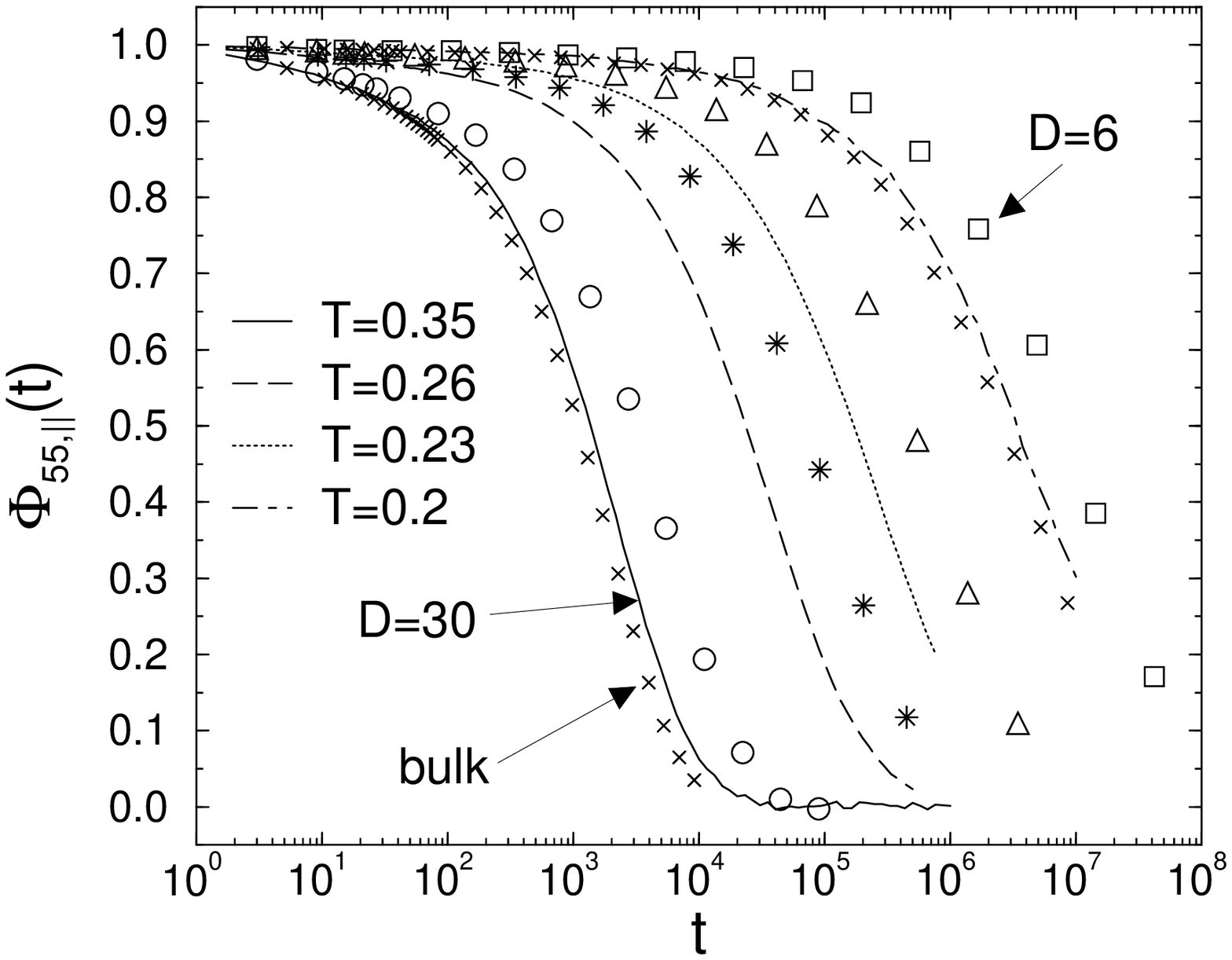} 
\vspace{4mm}
\caption[]{}
\end{figure}
\vspace*{15mm}
\begin{figure}
\epsfysize=110mm
\hspace*{-15mm}\epsffile{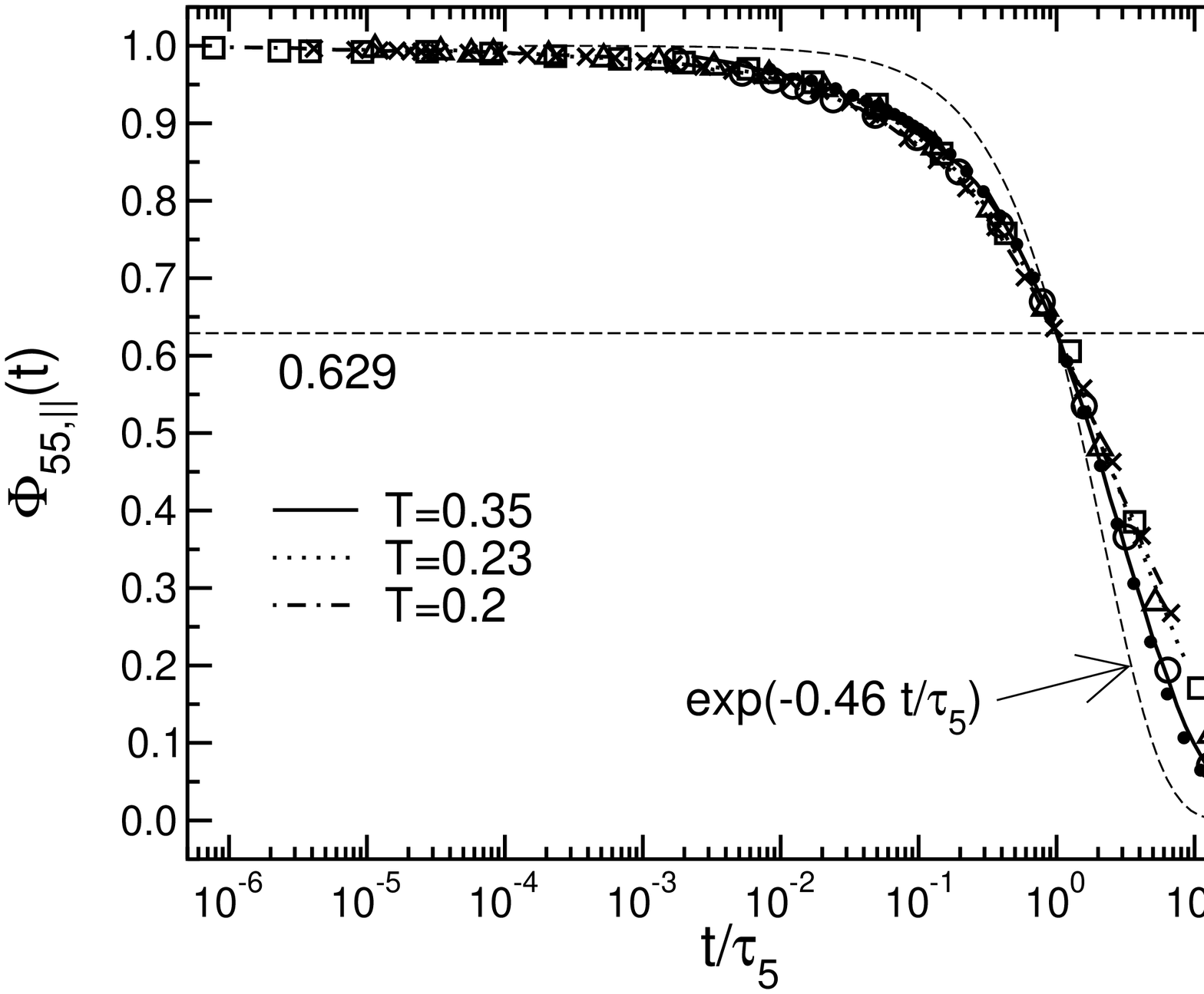} 
\vspace{10mm}
\caption[]{}
\end{figure}
\begin{figure}
\epsfysize=140mm
\hspace*{-20mm}\epsffile{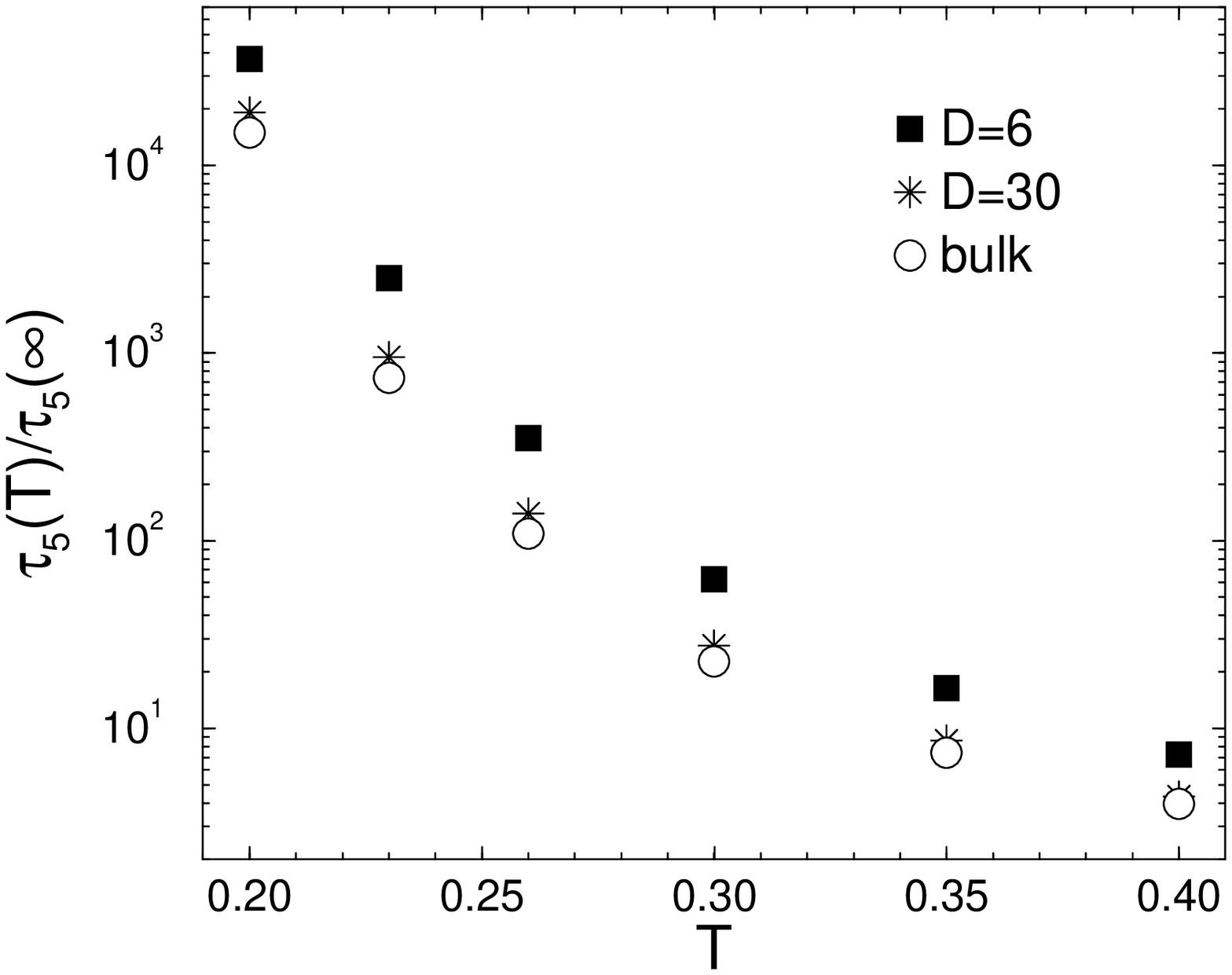} 
\vspace{4mm}
\caption[]{}
\end{figure}
\begin{figure}
\epsfysize=90mm
\hspace*{-36mm}\epsffile{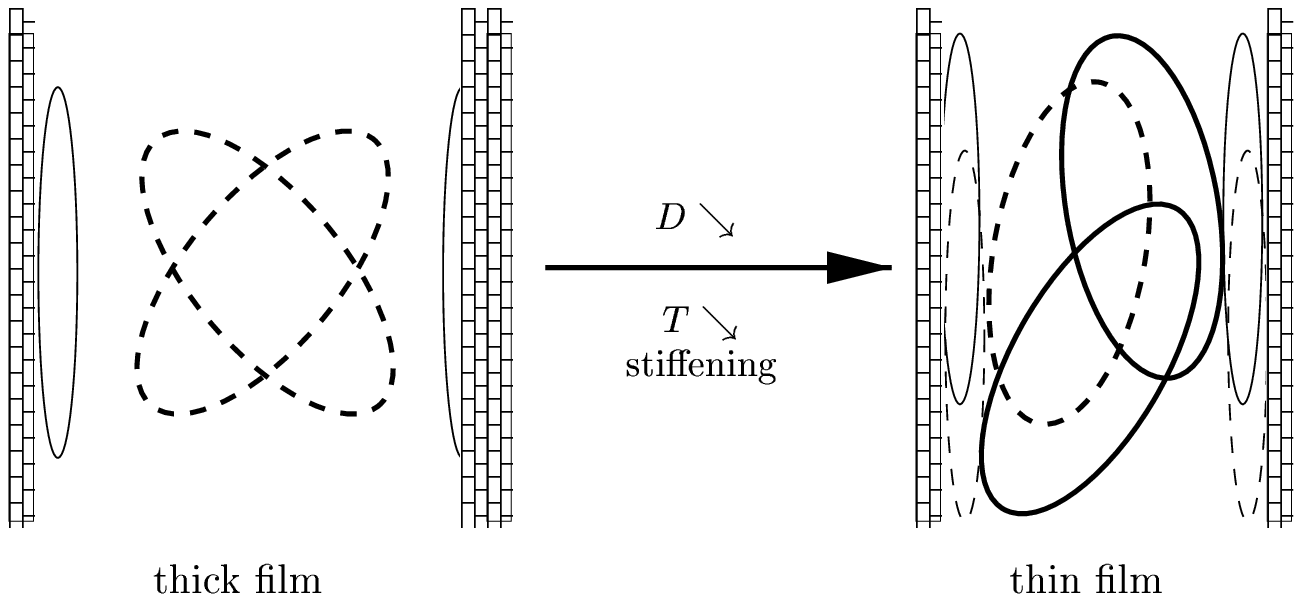}
\vspace{10mm}
\caption[]{}
\end{figure}
\newpage
\begin{figure}
\epsfysize=70mm
\hspace*{-36mm}\epsffile{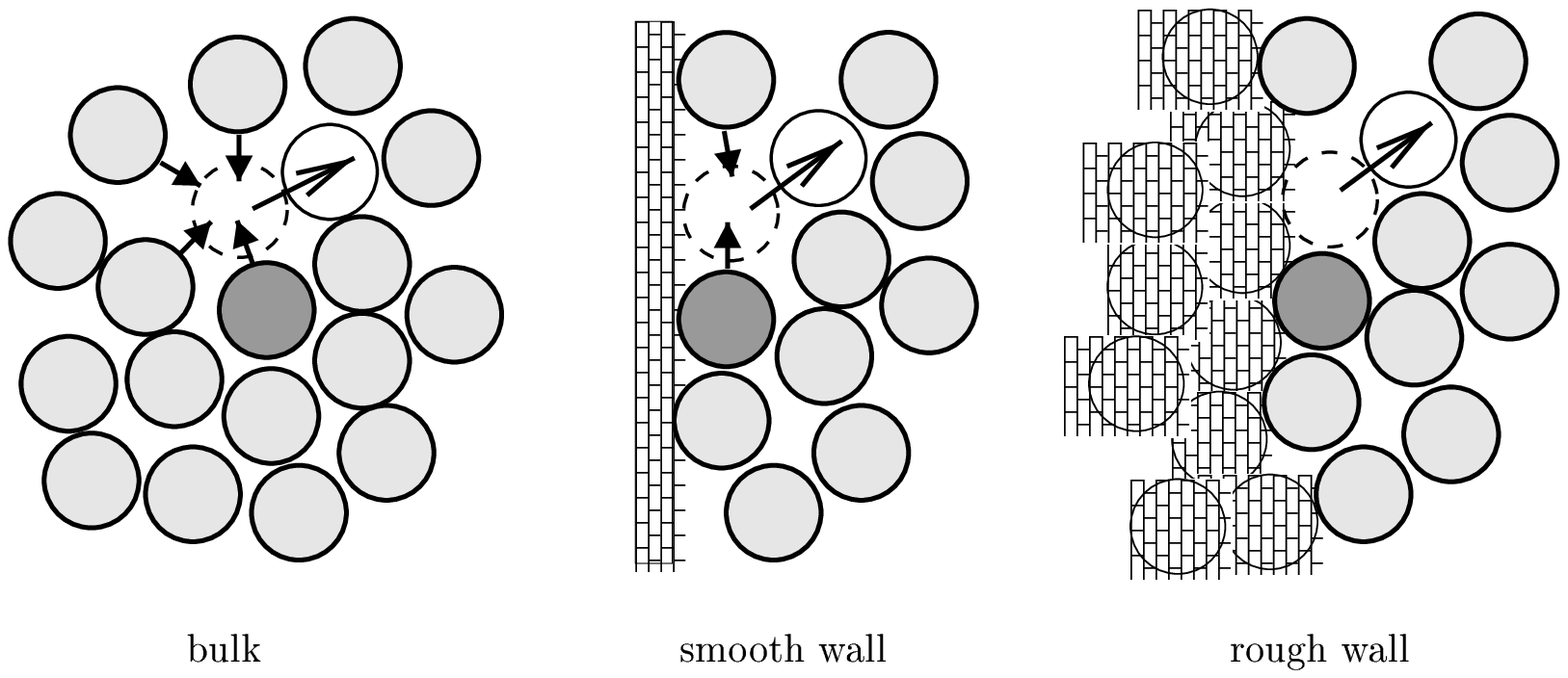}
\vspace{20mm}
\caption[]{}
\end{figure}
\end{document}